\documentclass[prc,showpacs,showkeys,superscriptaddress,nofootinbib]{revtex4}
\usepackage{dsfont}
\usepackage{graphicx,amsmath}
\def\be{\begin{eqnarray}}
\def\ee{\end{eqnarray}}
\def\bc{\begin{center}}
\def\ec{\end{center}}
\def\dsp{\displaystyle}
\def\txst{\textstyle}
\def\rmF{{\rm F}}
\def\rmd{{\rm d}}
\def\om{\omega}
\def\scr{\scriptstyle}
\newcommand{\lsim}{\stackrel{\scriptstyle <}{\phantom{}_{\sim}}}

\def\piproj{\hat{\pi}}
\begin{document}
\title{Neutral weak currents in nucleon superfluid Fermi liquids:
\\
Larkin-Migdal and Leggett approaches}
\author{E.E. Kolomeitsev}
\affiliation{Matej Bel  University, SK-97401 Banska Bystrica, Slovakia}
\author{D.N. Voskresensky}
\affiliation{Moscow Engineering Physics Institute (MEPhI),\\ Kashirskoe
  Avenue 31, RU-115409 Moscow, Russia}
\affiliation{GSI, Plankstra\ss{}e 1, D-64291 Darmstadt, Germany}
\begin{abstract}
Neutrino emission in processes of breaking and formation of
nucleon Cooper pairs is calculated in the framework of the
Larkin-Migdal and the Leggett approaches to the description of
superfluid Fermi liquids at finite temperatures. We explain
peculiarities of both approaches and explicitly demonstrate that
they lead to the same expression for the emissivity in pair
breaking and formation processes.
\end{abstract}
\date{\today}
\maketitle

\section{Introduction}

One of important mechanisms for  cooling of superfluid neutron star interiors  is  nucleon Cooper
pair breaking and formation (PBF) with a radiation of neutrino-antineutrino pairs, $"N"\to "N"
+\nu+\bar{\nu}$ (see
\cite{FRS76,VS87,SV87,MSTV90,SVSWW97,Minimal,KHY,YKL99,YLS99,V01,BGV04,GV05,PGW,Sedr07,KR,LP,SMS,KV08,L08,SR09,LeinsonWrong}
and references therein). Neutrinos are produced in weak interactions in which the lepton current
$l_\mu=\bar{\nu}(1-\gamma_5)\nu$ is coupled to the vector ($V_\mu=g_V\,\bar{\Psi}\, \gamma^\mu\,
\Psi$) and the axial-vector ($A_\mu= g_A\,\bar{\Psi}\, \gamma^\mu \gamma_5\, \Psi$) nucleon
currents; $\mathcal{L}=-\frac{G}{2\,\sqrt 2}\,(V^\mu-A^\mu)\,l_\mu$, where $G\approx 1.2\times
10^{-5}$~GeV$^{-2}$ is the weak interaction coupling, $g_V=-1$ and $g_A=1.26$. For the nuclear
vector current holds $\partial_\mu V^\mu=0$ that corresponds to the conservation of the baryon
charge. Early
works~\cite{FRS76,VS87,SV87,MSTV90,SVSWW97,Minimal,KHY,YKL99,YLS99,V01,BGV04,GV05,PGW,Sedr07} did
not care about the vector-current conservation. The latter is fulfilled only if  in-medium
renormalization of the vector current is performed  together with a corresponding renormalization
of Green's functions. This problem was tackled in Refs.~\cite{KR,LP,SMS,KV08,L08,SR09,PLPS}.
Reference~\cite{LP} indicated that the emissivity of the $1S_0$ PBF processes on the vector
current should be dramatically suppressed ($\propto v_{\rm F}^4$, where $v_{\rm F}$ is the Fermi
velocity of non-relativistic nucleons) provided the vector current conservation is fulfilled.
Reference~\cite{LP} used expressions derived within the standard BCS formalism of the
superconductivity theory \cite{Nambu,Schriffer} for low excitation energies $\om, |\vec{q}|\ll
\Delta$, where $\Delta$ is the nucleon pairing gap and $\om$ is the net $\nu\bar{\nu}$ energy,
whereas the PBF reaction kinematics permits only $\om >2\Delta$, $|\vec{q}|<\om$. The correlation
effects in the particle-hole channel were neglected and processes induced by the axial-vector
current were disregarded.

The consistent calculation of the PBF emissivity induced by the vector and axial-vector currents
was performed in Ref.~\cite{KV08} within the Larkin-Migdal-Leggett Fermi liquid approach. The
latter takes properly into account of the correlation effects in both particle-particle and
particle-hole channels. It was demonstrated that the neutrino emissivity is actually controlled by
the axial-vector current and is suppressed only by the factor $\propto v_{\rm F}^2$, rather than
$\propto v_{\rm F}^4$ as was stated in Ref.~\cite{LP}. This result was supported in the subsequent
work~\cite{L08}, which however continued to work out the vector current contribution neglecting
the correlation effects in the particle-hole channel.

The convenient Nambu-Gorkov formalism developed for the description of metallic superconductors,
cf. Refs.~\cite{Nambu,Schriffer}, does not distinguish interactions in particle-particle and
particle-hole channels. These interactions can be, however, essentially different in strongly
interacting matter, like in nuclear matter and in liquid He$^3$. The adequate methods for Fermi
liquids with pairing were developed for zero temperature by Larkin and Migdal in Ref.~\cite{LM63}
(see also \cite{M67}) and for a finite temperature by Leggett in Ref.~\cite{Leg65a,L66}. The
problem of calculation of a response function of a Fermi system to an external interaction becomes
tractable at cost of introduction of a set of Landau-Migdal parameters for quasiparticle
interactions. Parameters can be either evaluated microscopically or extracted from analysis of
experimental data, see \cite{M67}. The technical difference of the mentioned approaches is that
Larkin and Migdal worked out equations for full in-medium vertices, whereas Leggett calculated
directly a response function. The former approach was aimed at the study of transitions in nuclei,
and the latter on the analyzes of collective modes in superfluid Fermi liquid. The principal
equivalence of both approaches was emphasized by Leggett in Ref.~\cite{Leg65a,L66}.

In Ref.~\cite{KV08} we used the Larkin-Migdal approach. More specifically, we solved the
Larkin-Migdal equations for the vertices induced by the weak vector and axial-vector currents and
calculated the current-current correlation function $\chi_{a}$, $a =\{V,A\}$, and the PBF
emissivity. The explicit expression for $\chi_{a}$ was obtained at $T=0$. It is sufficient to
calculate emissivity of the process for $T\ll 2\Delta$ since small exponential factor
$e^{-2\Delta/T}$ comes already from the phase space volume and the temperature correction to
$\chi_{a}(T=0)$ is small in this limit. Despite Larkin-Migdal equations were derived in their
original paper for $T=0$, actually the results can be generalized for arbitrary $T$. In
Ref.~\cite{KV08} we sketched  which expressions should be modified at finite temperatures.

Recent paper~\cite{LeinsonWrong} tried to adopt the Leggett formalism to calculate the PBF
emissivity for the $1S_0$ neutron pairing and came to different results compared to those derived
in Ref.~\cite{KV08}, even in case $T\ll \Delta$. One of the points of Ref.~\cite{LeinsonWrong} was to
find the PBF emissivity for arbitrary $T\neq 0$. In view of the explicit claim by Leggett
in Ref.~\cite{L66} on the equivalence of his approach to that of Larkin and Migdal, this difference
looks worrisome and requires a clarification.

The aim of this paper is to reveal the correspondence between the Larkin-Migdal and Leggett
approaches and to generalize the results of Ref.~\cite{KV08} to arbitrary temperatures. We argue
also that the results of the work \cite{LeinsonWrong} are based on misinterpretation and wrong
solution of the Leggett equations.

The paper is organized as follows. In Section~\ref{sec:KV} we introduce the Fermi liquid approach
to the problem of the neutrino emissivity from nucleon matter. We focus on the emissivity via the
PBF reactions. The main expressions are presented in the diagrammatic form valid at arbitrary
temperatures. Then in Section~\ref{sec:Correspondence} we demonstrate how the Larkin-Migdal
equations and the Leggett equations follow from the same set of diagrams. In
Section~\ref{sec:Solution} we solve the Larkin-Migdal equations for the vertices induced by the
vector and axial-vector currents and, then, present expressions for the PBF emissivity recovering
the results of Ref.~\cite{KV08} formulated now for arbitrary temperatures. In
Section~\ref{sec:Leggett} we demonstrate how the same results can be obtained within the Leggett
formalism. In Section~\ref{sec:Critic} we discuss the flaws in Ref.~\cite{LeinsonWrong}. We
conclude with Section~\ref{sec:Conclude}.

\section{General expressions for neutrino emissivity and  current-current correlators}
\label{sec:KV}

In this section we recall the main expressions for the neutrino emissivity obtained in
Ref.~\cite{KV08}, writing them in the form valid at arbitrary temperatures. To be specific we
consider a neutron system with the s-wave pairing. The neutrino emissivity
\be
\varepsilon_{\nu\bar\nu}=\frac{G^2}{4}
\intop \prod_{i=1}^2\frac{d^3 {q}_i}{(2\pi)^3\,2\, \om_i}
\frac{(\om_1+\om_2)\,\Im \sum\chi(q_1+q_2)}{e^{(\om_1+\om_2)/T}-1}
\label{emissivity}
\ee
is expressed through the imaginary part of
the Fourier-transform of the current-current correlator
$\chi(q)=\chi_V (q)+\chi_A (q)= \int\rmd^4 x e^{-i\, (q\cdot x)}
\langle N|\big(V^\mu(x)\, V^\nu(0)+A^\mu(x)\,A^\nu(0)\big)\,l_\mu(x)l_\nu^\dag(0)|N\rangle$\,, where
$q=(\om_1 +\om_2, \vec{q}_1 +\vec{q}_2)$.
The averaging is done over the vector of state, $|N\rangle$, of the fermion system with pairing at
thermal equilibrium. The summation runs over the lepton spins.

We use non-relativistic limit for nucleons  since the nucleon Fermi energy $\epsilon_{\rmF}\ll m$,
where $m$ is the nucleon mass. The bare vertices for the nucleon currents as they follow from the
expansion of the Lagrangian are
$V^{0\mu}\approx\psi^\dag\big({p}')\,(1,\,(\vec{p\,}'+\vec{p\,})/2\,m\big)\, \psi (p)$ and
$A^{0\mu}\approx \psi^\dag(p')\,\big(\vec{\sigma}(\vec{p\,}'+\vec{p\,})/2\,m
,\,\vec{\sigma}\big)\, \psi(p)$\,, where $\vec{\sigma}=(\sigma_1,\sigma_2,\sigma_3)$ are the Pauli
matrices acting on free nucleon spinors $\psi$, and $\vec{p\,}'$ and $\vec{p\,}$ are outgoing and
incoming momenta.

\begin{figure}
\includegraphics[width=8.2cm]{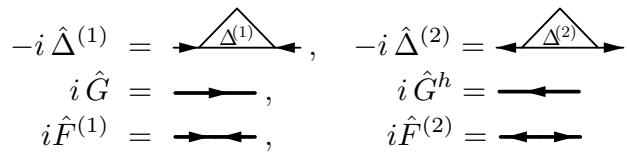}
\caption{ \label{fig:Delta} Amplitudes of the transition of a
particle into a hole and a condensate pair, $\Delta^{(1)}$ and vice versa, $\Delta^{(2)}$, and
normal, $G$, and anomalous, $F$, Green's functions.}\end{figure}

In a system with pairing a particle can  transit into a hole and
a condensate pair and vice versa. The one-particle one-hole irreducible amplitudes
of such processes are depicted in Fig.~\ref{fig:Delta}.
Besides the normal Green's function $\,\hat{G} $\, one introduces anomalous Green's functions
$\hat{F}^{(1)}$\,, and $\hat{F}^{(2)}$\,, which diagrammatic notations are given in
Fig.~\ref{fig:Delta}. The Green's function for the hole is defined as $\hat{G}^h(p)=\hat{G}^{\rm T}(-p)$\,.
Superscript $\rm T$ denotes matrix
transposition. The normal
and anomalous Green's functions are related by Gor'kov equations. The normal Green's function is
diagonal in spin space $\hat{G}(p)=G(p)\,\hat{1}$, but the anomalous Green's functions are equal
to $\hat{F}^{(1)}(p)=\hat{F}^{(2)}(p)=\hat{F}(p)=F(p)\, i\sigma_2$. We will not discuss here the
details of the Fermi liquid renormalization procedure, see Refs.~\cite{Migdal63,LM63,Leg65a,M67}.
The latter allows to separate contributions from the pole and regular parts of the  Green's
functions in all relevant quantities. We  assume that this procedure is properly done and we deal
further only with the pole parts of the Green's functions characterized by the effective mass
$m^*$ and the residue $a$:
 \be
G(p)= \frac{a\,(\epsilon +\epsilon_p)}{\epsilon^2-E_p^2 +i0{\rm
sgn}\epsilon },\, F(p)=\frac{-a\, \Delta}{\epsilon^2-E_p^2 +i0{\rm
sgn}\epsilon }\,, \label{GFpole}
 \ee
 where
$E_p^2=\epsilon_p^2+\Delta^2$ with the pairing gap $\Delta$, and $\epsilon_p=p^2/(2\, m^*)-\mu
(T)$, $\mu$ is the chemical potential,  $\mu (T=0)= \epsilon_\rmF=p_\rmF^2/(2\, m^*)$. At finite
temperature one can use the Matsubara techniques with the replacement
 $\epsilon\to i\,\epsilon_n =i\,\pi (2n+1)$\,.

The amplitude in the particle-particle channel is parameterized as
\be
\big[\widehat{\Gamma}^\xi\big]^{ac}_{bd}=
\Gamma_0^\xi(\vec{n},\vec{n\,}')(i\sigma_2)^a_b (i\sigma_2)^c_d
+ \Gamma^\xi_1(\vec{n},\vec{n\,}') (i\sigma_2\,\vec{\sigma})^a_b (\vec{\sigma}\,i\sigma_2)^c_d
,\nonumber\ee
and the interaction in the particle-hole channel is
\be
\big[\widehat{\Gamma}^\om\big]^{ac}_{bd}=
\Gamma_0^\om(\vec{n},\vec{n\,}')\,\delta^a_b\,
\delta^c_d + \Gamma^\om_1(\vec{n},\vec{n\,}')\, (\vec{\sigma})^a_b\,
(\vec{\sigma})^c_d .
\nonumber
\ee

Here and below $\vec{n}=\vec{p}/|\vec{p\,}|$ and
$\vec{n\,}'=\vec{p\,}'/|\vec{p\,}'|$\,. Superscript "$\om$"
indicates that the amplitude is taken for
$|\vec{q\,}\vec{v\,}_{{\rm F}}|\ll \om$ and $\om\ll \epsilon_{{\rm
F}}$, where $\om$ and $\vec{q}$ are transferred  energy and
momentum. We use that $\epsilon_{\vec{p}+\vec{q}/2}\approx
\epsilon_p+\vec{v}\,\vec{q}/2$, where $\vec{v}$ is the nucleon
velocity at the Fermi surface, $\vec{v}=v_\rmF\,
\vec{n}(1+O(T^2/\epsilon_{\rm F}^2))$. Since in the PBF processes
$|\vec{q}\,|\lsim \Delta$, $\om \sim 2\Delta$, the terms neglected
are of the order of $O(\Delta^2/p_{\rm F}^2)$. Actually, the
denominators of the Green's functions are $|\om \pm
(\epsilon_{\vec{p}+\vec{q}/2}-\epsilon_{\vec{p}-\vec{q}/2})|\ll
\epsilon_{{\rm F}}$. Moreover, the terms $\propto\vec{v}\,\vec{q}$
may vanish under the angular integrations. Taking this into
account we estimate that the neglected terms are at most of the
order of $\Delta/\epsilon_{\rm F}\ll 1$ compared to the retained
terms. Such corrections are usually omitted in most calculations
within the Fermi liquid theory for superfluids.

The empirical information is usually expressed in terms of
dimensionless parameters
 \be
\Gamma_0^{\om,\xi}(\vec{n},\vec{n\,}')= \frac{f^{\om,\xi}(\vec{n},\vec{n\,}')}{a^2\rho(n_0)}\,, \,\,
\Gamma_1^{\om,\xi}(\vec{n},\vec{n\,}')= \frac{g^{\om,\xi}(\vec{n},\vec{n\,}')}{a^2\rho(n_0)}\,,
\label{LMP}
 \ee
where $\rho=\frac{m^*\,p_{{\rmF}}}{\pi^2}$ is the density of states at the Fermi surface.

The pairing gap is determined by the $\Gamma_{0}^\xi$ term in the particle-particle interaction,
and the gap equation reads
 \be
&&\Delta(\vec{n}) = - \langle \Gamma_0^\xi(\vec{n},\vec{n}')
\,A_0(\Delta(\vec{n\,}')) \, \Delta(\vec{n\,}')
\rangle_{\vec{n}'}\,,
\label{gapeq}\\
&&A_0(\Delta)= \int
\rmd\Phi_T G_0(p)\, G^h (p)\, \theta(\xi-\epsilon_p), \label{A0}
 \ee
where $G_0(p)=1/(\epsilon -\epsilon_p+i0{\rm sgn}\epsilon)$ is the Green's function for the Fermi
system without pairing ($\Delta=0$),  and $\xi\sim \mu$ is the cutting parameter. Usually the gap
is determined with $\xi =\mu$. We use the following notations for the angular integration
 \be
\langle \dots\rangle_{\vec{n}}&=&\int\frac{\rmd \Omega_{\vec n}}{4\pi}\, (\dots)\,,
 \ee
and for the integration
 \be
&& \int \rmd\Phi_T \,f(\epsilon,\epsilon_p) =\left\{
\begin{array}{cc}
\dsp \rho\,\int_{-\infty}^{+\infty}
\frac{\rmd \epsilon}{2\pi i}
\int_{-\infty}^{+\infty}\rmd\epsilon_p f(\epsilon,\epsilon_p)
& \mbox{for}~T=0\\
\dsp \rho \,T \sum_{n=-\infty}^{\infty}
\int_{-\infty}^{+\infty}\rmd\epsilon_p f(i\epsilon_n,\epsilon_p)
& \mbox{for}~T\neq 0
\end{array}\right.\,.
\nonumber
 \ee
Note that the same value $A_0$ can be introduced as $A_0= \int
\rmd \Phi_T\,(G(p)\, G^h(p)+F(p)\, F(p))\,
\theta(\xi-\epsilon_p)$, cf. Ref.~\cite{L66}.\footnote{Definition
of the value $A_0$ is here the same as in Ref. \cite{LM63}  and
differs by sign from that used in Ref.~\cite{Leg65a,L66}.}

\begin{figure}
\parbox{8.2cm}{\includegraphics[width=8.2cm]{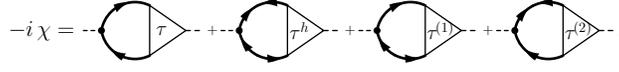}}
\caption{\label{fig:chi} Diagrams contributing to the
susceptibility $\chi$ in (\ref{emissivity}).}
\end{figure}
The diagrammatic representation for the  current-current correlator entering (\ref{emissivity}) is
shown in Fig.~\ref{fig:chi}. The dash line relates to the $Z$-boson coupled to the neutral lepton
currents. In terms of the Green's functions and the quasiparticle interactions we write it as
 \be
&&\chi_a(\om,\vec{q}\,)= \frac{1}{2}{\rm Tr}
\Big\langle
\int d\Phi_T\hat{\tau}_a^\om\,
\Big\{
 \hat{G}_{+}\,\hat{\tau}_a\,\hat{G}_{-}
+\hat{F}_{+}\,\hat{\tau}_a^{h}\,\hat{F}_{-}
+\hat{G}_{+}\,\hat{\tau}_a^{(1)}\hat{F}_{-}
+\hat{F}_{+}\,\hat{\tau}_a^{(2)}\hat{G}_{-} \Big\}
\Big\rangle_{\vec{n}}\,, \label{chi:form}
 \ee
$a=V$ for the vector current and  $a=A$ for the axial-vector currents. The trace is taken in the
nucleon spin space. Doing calculations in Matsubara technique, we use  notations:
$G_{+}=G(p_0+\om,\vec{p}+\vec{q}/2)$ and $G_{-}=G(p_0,\vec{p}-\vec{q}/2)$, and similarly for
$F_\pm$ functions. For continues frequencies we use the symmetrical 4-vector notations, i.e.
$G_\pm=G(p\pm q/2)$ and, analogously, $F_\pm =F(p\pm q/2)$. The left vertices in Fig.~\ref{fig:chi}
are the bare vertices generated by the weak currents
 \begin{subequations}
 \label{tauom}
 \be
\hat\tau^\om_V(\vec{n},q) &=& g_V\, \big(\tau^\om_{V,0}\,l_0-\vec{\tau}^\om_{V,1}\, \vec{l}\,\,\big)\hat{1}\, ,
\quad
\tau^\om_{V,0}=\frac{e_V}{a}\,\,,
\quad
\vec{\tau\,}^\om_{V,1}=\frac{e_V}{a}\, \vec{v},
\label{VC:tauom}\\
\hat\tau^\om_A(\vec{n},q) &=& -g_A\,\big(\vec{\tau\,}^\om_{A,1}\,\vec{\sigma}\,l_0
                           -\tau^\om_{A,0}\, \vec{\sigma}\vec{l}\,\,\big)\, ,
\quad
\tau^\om_{A,0}=\frac{e_A}{a}\,,
\quad \vec{\tau\,}^\om_{A,1}=\frac{e_A}{a}\, \vec{v}\,,
\label{AC:tauom}
 \ee
\end{subequations}
$e_V$ and $e_A$ are effective charges of the vector and
axial-vector currents, cf.~\cite{KV08}. The corresponding vertices
for holes are  defined as
\begin{subequations}
\label{tauomh}
\be
\hat\tau^{\om,h}_V(\vec{n},q) &=& [\hat\tau^{\om}_V(-\vec{n},q)]^{\rm T}
    =g_V\, \big(\tau^\om_{V,0}\,l_0+\vec{\tau\,}^\om_{V,1}\, \vec{l}\,\,\big)\hat{1},
\label{VC:tauomh}\\
\hat\tau^{\om,h}_A(\vec{n},q) &=&  [\hat\tau^{\om}_A(-\vec{n},q)]^{\rm T}
    =-g_A\, \big(-\vec{\tau\,}^\om_{A,1}\,\vec{\sigma\,}^{\rm T} \,l_0-
           \tau^\om_{A,0}\, \vec{\sigma\,}^{\rm T}\vec{l}\,\,\big)\,.
\label{AC:tauomh}
 \ee
 \end{subequations}
We used here explicitly that $\tau^\om_{a,0}(-p,q)=\tau^\om_{a,0}(p,q)$ and
$\vec{\tau\,}^\om_{a,1}(-p,q)=-\vec{\tau\,}^\om_{a,1}(p,q)$.

\begin{figure*}
\centerline{\includegraphics[width=17cm]{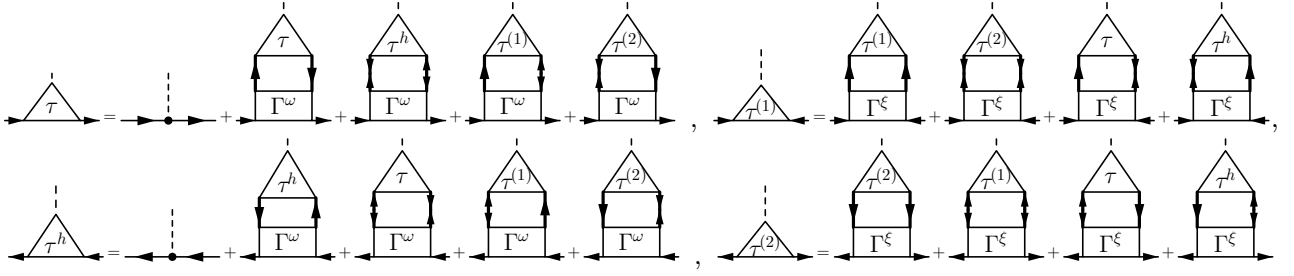}}
\caption{Graphical representation of dressed vertices in
Fig.~\ref{fig:chi}. \label{fig:verteq}}
\end{figure*}

The right vertices in Fig.~(\ref{fig:chi}) are the full in-medium-dressed vertices, which are
functions of the out-going frequency $\om$ and momentum $\vec{q}$, and
the nucleon velocity $\vec{v}\simeq v_\rmF\, \vec{n}$,
$\vec{n}=\vec{p}/p$\,. In absence of external magnetic field the spin structure of the full vertex
is the same as the spin structure of the bare vertex. Therefore, we write
 \begin{subequations}
\label{full}
 \be
\hat{\tau}_V&=&g_V\, \big(\tau_{V,0}\, l_0-\vec{\tau}_{V,1}\,\vec{l\,}\big)\,\hat{1}\,, \quad
\hat{\tau}^h_V=g_V\, \big(\tau_{V,0}^h\, l_0-\vec{\tau\,}_{V,1}^h\,\vec{l}\,\,\big)\,\hat{1}\,,
\label{full:tv}\\
\hat{\tau}_V^{(1)}&=&(\hat{\tau}_V^{(2)})^\dag=
-g_V\,\big(\widetilde{\tau}_{V,0}\, l_0-\vec{\widetilde{\tau}}_{V,1}\,\vec{l}\,\,\big)\, i\, \sigma_2\,,
\label{full:ttv}\\
\hat{\tau}_A&=&-g_A\, \big(\vec{\tau}_{A,1}\vec{\sigma}\, l_0-\tau_{A,0}\,\vec{\sigma}\vec{l}\,\,\big)\,,
\quad
\hat{\tau}^h_A=-g_A\, \big(\vec{\tau\,}_{A,1}^h\vec{\sigma}^{\rm T}\, l_0 -
\tau_{A,0}^h\,\vec{\sigma\,}^{\rm T}\vec{l}\,\,\big)\,,
\label{full:ta}\\
\hat{\tau}^{(1)}_A &=&(\hat{\tau}^{(2)}_A)^\dag =+ g_A\, \big(\vec{\widetilde\tau}_{A,1}\,
\vec{\sigma}\, l_0-\widetilde{\tau}_{A,0}\, \vec{\sigma}\,
\vec{l}\,\,\big)\, i\, \sigma_2\,.
\label{full:tta}
 \ee
\end{subequations}
We anticipate here the relation between the vertices $\hat{\tau}^{(1)}$ and  $\hat{\tau}^{(2)}$,
which will be proven later. The full vertices are determined by the diagramatic equations depicted
in Fig.~\ref{fig:verteq}.
These graphical equations were first introduced by Larkin and Migdal \cite{LM63}. The blocks in
Fig.~\ref{fig:verteq} correspond to the two-particle irreducible interaction in the
particle-particle channel, $\Gamma^\xi$, and the particle-hole irreducible interaction  in the
particle-hole channel, $\Gamma^\om$. We emphasize that only chains of bubble diagrams are summed
up in this particular formulation. Thus, the imaginary part of $\chi$ accounts only for
one-nucleon processes. To include two-nucleon processes within a quasi-particle approximation,
one should add diagrams with self-energy insertions to the Green's functions and iterate
the Landau-Migdal amplitudes $\Gamma^{\om,\xi}$ in Fig.~\ref{fig:verteq} in the horizontal
channel~\cite{KV95}.

Working out the spin structure in (\ref{chi:form}) we obtain
 \begin{subequations}
\be
&&\chi_V(q)= g_V^2\,
\big\langle\big(\tau_{V,0}^\om\,l_0-\vec{\tau\,}^\om_{V,1}\, \vec{l}\,\big)\,
\big(l_0\, \chi_{V,0}(\vec{n},q) - \vec{\chi}_{V,1}(\vec{n},q)\, \vec{l}\,\,\big)
\big\rangle_{\vec{n}},
\label{chiV}\\
&&\chi_A(q) = g_A^2\,
\big\langle\big(\vec{\tau\,}^\om_{A,1}\,l_0-\tau_{A,0}^\om\,\vec{l}\,\big)\, \big(l_0\,
\vec{\chi}_{A,1}(\vec{n},q) - \chi_{A,0}(\vec{n},q)
\vec{l}\,\,\big) \big\rangle_{\vec{n}},
\label{chiA}
\ee
 \end{subequations}
where we introduced new scalar and vector response functions
 \begin{subequations}
\be
&&\chi_{a,0}(\vec{n},q) =
\intop \rmd \Phi_T \big(G_+\, G_-\, \tau_{a,0} - F_+\, F_- \,\tau^h_{a,0}
+(G_+\, F_- - F_+\, G_-)\,\widetilde{\tau}_{a,0}\big)\,,
\label{chi0}\\
&&\vec{\chi}_{a,1}(\vec{n},q)=\intop \rmd \Phi_T \big(G_+\, G_-\, \vec{\tau}_{a,1} +
F_+\, F_- \,\vec{\tau\,}^h_{a,1}
+(G_+\, F_- - F_+\, G_-)\,\widetilde{\vec{\tau}}_{a,0}\big)\,.
\label{chi1}
\ee
 \end{subequations}

It is instructive to express susceptibilities (\ref{chiV}) and (\ref{chiA}) in terms of the polarization
tensors
 \be
\chi_a(q)=l_{\mu}\,\Pi_a^{\mu\nu}(q)\, l_{\nu}\,. \nonumber
 \ee
The polarization tensors can be written through the auxiliary
bare and dressed currents, $\hat{J}_{a,\mu}^\om (\vec{n},q)$ and
$\hat{J}_{a,\mu}(\vec{n},q)$, respectively,
 \be
\Pi_{a}^{\mu\nu}(q)=\frac12\langle {\rm Tr}\{\hat{J}_{a}^{\om,\mu}
(\vec{n},q)\, \hat{J}_{a}^\nu(\vec{n},q)\} \rangle_{\vec{n}}\,,
\label{def:Pi}
 \ee
 where the vector currents are
\be
\hat{J}_{V}^{\om,\mu} (\vec{n},q)=\big(\tau_{V,0}^\om ,
\vec{\tau\,}^\om_{V,1}\big)\, \hat{1}\,, \quad \hat{J}_{V}^{\mu}
(\vec{n},q)=\big(\chi_{V,0}, \vec{\chi}_{V,1}\big)\, \hat{1}\,,
\label{JV}
 \ee
and  the axial-vector currents are
\be
\hat{J}_{A}^{\om,\mu} (\vec{n},q)=\big(\vec{\sigma\,}\vec{\tau\,}^\om_{A,1},
\vec{\sigma\,}\tau_{A,0}^\om \big)\,,
\quad
\hat{J}_{A}^{\mu} (\vec{n},q)=\big(\vec{\sigma\,}\vec{\chi\,}_{A,1}, \vec{\sigma\,}\chi_{A,0}\big)\,.
\label{JA}
 \ee
Thus, the polarization tensors can be cast as
\be
\Pi_V^{\mu\nu} &=&
\left(
\begin{array}{cc}
\Pi_V^{00}, & \Pi_V^{0j}\\ \Pi_V^{i0}, & \Pi_V^{ij}
\end{array}
\right)=\left(
\begin{array}{cc}
\langle\tau^\om_{V,0}\, \chi_{V,0}\rangle_{\vec{n}}, &
\langle\tau^\om_{V,0}\, (\vec{\chi\,}_{V,1})_j\rangle_{\vec{n}}\\
\langle(\vec{\tau\,}^{\om}_{V,1})_i\, \chi_{V,0}\rangle_{\vec{n}},
& \langle(\vec{\tau\,}^{\om}_{V,1})_i\,
(\vec{\chi}_{V,1})_j\rangle_{\vec{n}}
\end{array}
\right)\,, \label{PiV}\\ \Pi_A^{\mu\nu} &=&
\left(\begin{array}{cc} \Pi_A^{00}, & \Pi_A^{0j}  \\ \Pi_A^{i0}, &
\Pi_A^{ij}\\
\end{array}\right)
=
\left(\begin{array}{cl}
\langle\vec{\tau\,}^\om_{A,1}\vec{\chi}_{A,1}\rangle_{\vec{n}}, &
\langle\tau^\om_{A,0}\, (\vec{\chi\,\,}_{A,1})_j\rangle_{\vec{n}}
\\ \langle(\vec{\tau\,}^{\om}_{A,1})_i\,
\chi_{A,0}\rangle_{\vec{n}},&
\langle\tau^\om_{A,0}\,\chi_{A,0}\rangle_{\vec{n}}\,\delta^{ij}
\\
\end{array}\right)\,.
\label{PiA}
 \ee

The integration over the lepton phase space can be performed analytically, cf. \cite{KV08}, and
the neutrino emissivity is then  cast in terms of the polarization tensor as
 \be
\varepsilon_{\nu\nu} &=&
\varepsilon_{\nu\nu,V}+\varepsilon_{\nu\nu,A}\,, \nonumber\\
\varepsilon_{\nu\nu,a} &=& \frac{G^2}{48\, \pi^4}\,
\int^\infty_0\!\!\rmd \om\int_0^\om\!\!\rmd |\vec{q\,}|
\frac{\om\, \vec{q\,}^2}{e^{\om/T}-1}\, K_a(q)\,,
\label{emiss}\\
K_a(q) &=& \big(q_\mu\, q_\nu-g_{\mu\nu}\, q^2\big)\,\Im \Pi_a^{\mu\nu}(q)\, .
\label{kappa}
 \ee
The polarization tensor for the conserving vector
current must be transverse $\Pi_V^{\mu\nu}q_\nu=q_\mu\,
\Pi_V^{\mu\nu}=0$. This property will be explicitly proven below in Section~\ref{sec:Solution} and in
Appendix~\ref{app:Ward}.
Taking it into account and using (\ref{PiV},\ref{PiA}) we find
 \be
K_V(q)&=&
\frac{e_V\,g_V^2}{a} (\vec{q\,}^2-\om^2)\, \Im
\langle\chi_{V,0}(\vec{n},q) - \vec{v}\,
\vec{\chi}_{V,1}(\vec{n},q)\rangle_{\vec{n}}\,,
\label{kappaV:Ward}\\
K_A(q)&=&
\frac{e_A\,g_A^2}{a} \Im\big[
 \vec{q\,}^2 \langle \vec{v}\, \vec{\chi}_{A,1}(\vec{n},q)\rangle_{\vec{n}}
+ (3\,\om^2-2\, \vec{q\,}^2)\, \langle \chi_{A,0}(\vec{n},q)\rangle_{\vec{n}}
\nonumber\\
&-&\om \langle \vec{q}\, \vec{\chi}_{A,1}(\vec{n},q)\rangle_{\vec{n}}
-\om\,\langle(\vec{q}\,\vec{v}\,)\, \chi_{A,0}(\vec{n},q)\rangle_{\vec{n}}
\big]\,.
\label{kappaA}
 \ee

Finally, once the full in-medium vertices
(\ref{full:tv},\ref{full:ttv},\ref{full:ta},\ref{full:tta}) are known,  expressions
(\ref{emiss},\ref{kappaV:Ward},\ref{kappaA}) together with (\ref{chi0},\ref{chi1}) solve the
problem of neutrino emission from superfluid neutron matter via the PBF reactions.

In the rest of the paper we demonstrate two methods for the calculation of the full vertices.

\section{Correspondence between Larkin-Migdal and Leggett formalisms}\label{sec:Correspondence}

The coupling  of an external field to the non-relativistic fermion is described by the $2\times 2$
matrix acting in the fermion spin space. Any rank-2 matrix  can be decomposed into a unity matrix
and Pauli matrices $\vec{\sigma}$. Thus, we have
 \be
\hat{\tau}(\vec{n},q) &=& t_0(\vec{n},q)\, \hat{1} + \vec{\sigma}\,\vec{t}_1(\vec{n},q)\,,
\nonumber\\
\hat{\tau}^{(1)}(\vec{n},q)&=& \big(t_0^{(1)}(\vec{n},q)\, \hat{1}
         + \vec{\sigma}\,\vec{t}_1^{\,\,(1)}(\vec{n},q)\big)\, i\sigma_2,
\nonumber\\
\hat{\tau}^{(2)}(\vec{n},q)&=& i\sigma_2\,\big(t_0^{(2)}(\vec{n},q)\, \hat{1}
           + \vec{\sigma}\,\vec{t}_1^{\,\,(2)}(\vec{n},q)\big)\,.
\label{taut}
 \ee
The  hole-vertex is decomposed as
 \be
\hat{\tau}^h(\vec{n},q)=t_0^h(\vec{n},q)\, \hat{1} +
\vec{\sigma}^{\,\rm T}\vec{t}_1^{\,h}(\vec{n},q) =t_0(-\vec{n},q)\,
\hat{1} + \vec{\sigma}^{\,\rm T}\vec{t}(-\vec{n},q).
 \ee
The spin structures of the weak coupling vertices (\ref{tauom}) and (\ref{full}), we  primarily
deal with, demonstrate that the vector  current contributes only to the vertices with subscript 0
and the axial-vector  current couples only to the vertices with subscript 1. E.g., for the vector
current
 \be
t_0^{\om}  &=& g_V\, \big(\tau_{V,0}^\om \,l_0-\vec{\tau\,}_{V,1}^\om \, \vec{l\,}\big)\,,
\nonumber\\
t_0 &=& g_V\, \big(\tau_{V,0} \,l_0-\vec{\tau\,}_{V,1} \, \vec{l\,}\big)\,,
\nonumber\\
\vec{t\,}_1^{\om} &=& \vec{t}_1=0\,,
\label{VC:t0}
 \ee
and for the axial-vector current
 \be
t_0^\om &=& t_0=0\,,
\nonumber\\
\vec{t\,}_1^{\om}  &=& -g_A\, \big(\vec{\tau\,}_{A,1}^\om \,l_0- \tau_{A,0}^\om\,\vec{l\,}\big)\,,
\nonumber\\
\vec{t}_1  &=& -g_A\, \big(\vec{\tau\,}_{A,1} \,l_0- \tau_{A,0}\,\vec{l\,}\big)\,.
\label{AC:t1}
 \ee
After opening the spin structure of the diagrams in Fig.~\ref{fig:verteq} we arrive at the
following set of equations for $t_0$, $t_0^h$, $t_0^{(1)}$ and $t_0^{(2)}$ (for brevity we omit
the dependence of the vertices on $\vec{n}$, $\om$ and
$\vec{q\,}$):
 \begin{subequations}
\label{LManalytic}
\be
t_0 -t_0^\om &=&\phantom{-}
\Big\langle
\int \rmd\Phi_T\, \Gamma_0^\om\,
\Big[G_{+}\, G_{-}\, t_0 - F_{+}\, F_{-}\,t^h_0 - G_{+}\, F_{-}\, t^{(1)}_0 - F_{+}\, G_{-}\, t^{(2)}_0\Big]
\Big\rangle_{\vec{n\,}'}\,,
\label{LMan:t0}\\
t^h_0 -t^{h\om}_0 &=& \phantom{-}
\Big\langle
\int\rmd\Phi_T\,\Gamma_0^\om\,
\Big[G^h_+\, G^h_-\,t^h_0 - F_+\, F_-\, t_0 -F_+\,G^h_-\, t^{(1)}_0 - G_+^h\, F_-\,t^{(2)}_0 \Big]
\Big\rangle_{\vec{n\,}'}\,,
\label{LMan:th0}\\
t^{(1)}_0 &=& -\Big\langle
\int\rmd\Phi_T\,\Gamma_0^\xi\,
\Big[G_{+}\, G^h_{-}\,t^{(1)}_0 -F_{+}\,F_{-}\, t^{(2)}_0 +G_{+}\, F_{-}\, t_0 + F_{+}\, G^h_{-}\, t^h_0 \Big]
\Big\rangle_{\vec{n\,}'}\,,
\label{LMan:t10}\\
t^{(2)}_0 &=&-\Big\langle
\int\rmd\Phi_T\,\Gamma_0^\xi\,
\Big[G^h_+\,G_-\, t^{(2)}_0 - F_+\,F_-\,t^{(1)}_0 + F_+\, G_- \, t_0 + G^h_+\,F_-\, t^h_0\Big]
\Big\rangle_{\vec{n\,}'}\,.
\label{LMan:t20}
\ee
 \end{subequations}
The similar set of equations for 3-vector vertices $\vec{t}_1$, $\vec{t}_1^{\,h}$,
$\vec{t}_1^{\,(1)}$ and $\vec{t}_1^{\,(2)}$ is written with the only differences that
$\Gamma_0^{\om,\xi}$ is replaced by $\Gamma_1^{\om,\xi}$ and in front of all terms with $t^h$ the
sign must be changed. The origin of this sign change is the identity $\sigma_2\, \vec{\sigma}^{\rm
T}\, \sigma_2=-\vec{\sigma}$\,.

For the sake of convenience we introduce brief notations, e.g.,
 \be\int \rmd\Phi_T\,  G^{h}_{+}\,F_{-}=G^{h}_{+}\cdot F_{-}.
\label{dotoper}
 \ee
The details of calculations of these products within the  Matsubara techniques
 are deferred to Appendix~\ref{app:loop}. E.g., we recover useful relations
 \be
G_{+}\!\cdot\! G^h_{-}=   G^h_{+}\!\cdot\! G_{-}\,,\quad
G_{+}\!\cdot\! F_{-}  = - F_+\!\cdot\! G_-\,,\quad F_{+}\!\cdot\!
G^h_{-} =-  G^h_+ \!\cdot\! F_-\,, \label{relat1}
 \ee
  see Eqs.~(\ref{relation1}),
(\ref{relation2}) and (\ref{relation3}) in ~\cite{L66}.
From (\ref{LManalytic}) we can immediately find relations between the vertices $t^{(1)}_0$ and
$t^{(2)}_0$. Taking the sum of Eqs.~(\ref{LMan:t10}) and (\ref{LMan:t20}) and making use of
Eq.~(\ref{relat1}) we obtain the homogeneous equation for the sum $t^{(1)}_0+t^{(2)}_0$,
 \be
t^{(1)}_0+t^{(2)}_0 &=& -\Big\langle
\int\rmd\Phi_T\,\Gamma^\xi\,
\Big[G_{+}\, G^h_{-}\,-F_{+}\,F_{-}\Big]\, (t^{(1)}_0+t^{(2)}_0)
\Big\rangle_{\vec{n\,}'}\,,
\ee
which implies
\be
t^{(1)}_0+t^{(2)}_0=0\,.
\label{t12relat}
 \ee
The latter relation justifies the parameterization of the full in-medium vertices in
Eqs.~(\ref{full:ttv}) and (\ref{full:tta}). The same relation is valid for $\vec{t\,\,}^{(1)}_1$  and
$\vec{t\,\,}^{(2)}_1$ vertices.

\subsection{Larkin-Migdal equations}

In their original paper~\cite{LM63} Larkin and Migdal presented Eqs.~(\ref{LManalytic}) in
somewhat different form.  They noted that the vertices for the holes, $t^{h}$, can be obtained
from the particle vertices with the change $\vec{n}\to -\vec{n}$,
 \be
t_0^h(\vec{n},q)=t_0(-\vec{n},q)\,,
\quad
\vec{t\,}_1^h(\vec{n},q)=\vec{t}_1(-\vec{n},q)\,.
 \ee
Therefore, one can introduce the operator $\piproj$, which performs this change of $\vec{n}$ in the
vertex
 \be
\piproj\, t_0(\vec{n},q)=t_0^h(\vec{n},q)\,,\quad
\piproj\,\vec{t}_{1}(\vec{n},q)=\vec{t\,}_{1}^h(\vec{n},q)\,. \ee
In view of relation (\ref{t12relat}),
Eqs.~(\ref{LMan:t10},\ref{LMan:t20}) reduce to one equation for the vertex
 \be
\widetilde{t}_0=-t_0^{(1)}=t_0^{(2)}. \nonumber
 \ee
Analogously we introduce
$\widetilde{\vec{t}}_1=-\vec{t}_1^{\,(1)}=\vec{t\,}_1^{(2)}$\,.
Then  four Eqs.~(\ref{LManalytic}) for scalar vertices '0' and
four equations for 3-vector vertices $\vec{t}_1$ can be cast in
terms of four equations
\begin{subequations}
\label{LMeq}
\be
t_0-t_0^\om &=&
\Big\langle \Gamma_0^\om
\big( L(\piproj)\,t_0 +M\,\widetilde{t}_0\big)
\Big\rangle_{\vec{n\,}'}\, ,
\label{SwP:t0}
\\
\widetilde{t}_0 &=& -\Big\langle \Gamma_0^\xi\,
\Big( \big(N+A_0 \big)\,\widetilde{t}_0+O(\piproj )\,t_0 \Big)
\Big\rangle_{\vec{ n\,}'}
\label{SwP:tt0}\, ,
\\
\vec{t}_1 -\vec{t}^\om_1 &=&
\Big\langle \Gamma_{1}^{\om} \big( L(-\piproj )\,\vec{t}_1 +M\,\widetilde{\vec{t\,}}_1\big)
\Big\rangle_{\vec{n\,}'}
\label{SwP:t1}\, ,
\\
\widetilde{\vec{t}}_1 &=& -\Big\langle \Gamma^\xi_1\,
\Big(\big(N+A_0\big)\,\widetilde{\vec{t\,}}_1+O(-\piproj )\,\vec{t}_1 \Big)
\Big\rangle_{\vec{n\,}'}\,.
\label{SwP:tt1}
\ee
 \end{subequations}
We shall call this set of equations {\it the Larkin-Migdal equations}. In Ref.~\cite{LM63} these four
equations are further reduced to only two equations with the help of the operator $\hat{P}$
(defined by Eq.~(31) in Ref.~\cite{LM63}), which includes additionally the change of the sign between
Eqs.~(\ref{SwP:t0}) and (\ref{SwP:t1}) and between Eqs.~(\ref{SwP:tt0}) and (\ref{SwP:tt1}). Functions $L$,
$M$, $N$, and $O$ are defined as in Ref.~\cite{LM63}
 \be
L(\vec{n},q;\piproj ) &=&  G_+\!\cdot\! G_- - F_+\!\cdot\! F_-\,\piproj \,,
\nonumber\\
M(\vec{n},q)&=&  G_+\!\cdot\! F_- - F_+\!\cdot\! G_- \,,
\nonumber\\
N(\vec{n},q) &=& G_+\!\cdot\! G^h_- +F_+\!\cdot\! F_- - A_0\,,
\nonumber\\
O(\vec{n},q;\piproj ) &=& - G_+\!\cdot\! F_- - F_+\!\cdot\! G^h_- \piproj\, .
\label{depend}
 \ee
We emphasize that Eqs.~(\ref{LMeq}) are valid at arbitrary temperature. The temperature dependence
is hidden in the convolutions of the Green's functions (\ref{depend}). In Ref.~\cite{LM63} the
latter ones were calculated explicitly only for $T=0$ using the method of Ref.~\cite{VGL61}. The
extension to $T\neq 0$ is straightforward within the Matsubara technique, see Appendix~\ref{app:loop}.

\subsection{Leggett equations for vertices and correlation functions}

In this section, starting from Eqs.~(\ref{LManalytic}),  we derive the equations obtained by
Leggett in Refs.~\cite{Leg65a,L66}, thus, demonstrating interrelation of Larkin-Migdal
and Leggett approaches and their principal equivalence.

The primary goal of  works \cite{Leg65a,L66} was to calculate the response function of the
superfluid Fermi liquid to the source of an external field $\hat{\tau}^\om(\vec{n},q)$ at $\om,
q\ll \Delta$ and to study collective modes in a superfluid. The excitations of the density,
spin-density, current and spin current fluctuations with $\hat{\tau}^\om=\hat{1}$, $\sigma_3$,
$\vec{n}$ and $\vec{n}\, \sigma_3$, respectively, were considered. In Refs.~\cite{Leg65a,L66} it
was explicitly assumed that the source is diagonal in the spin space and the vertex is
characterized by the directions of $\vec{n}$ and  of the projection of the spin $s$. Then the
vertices can be classified as "even" and "odd" depending on whether they change the sign at the
simultaneous replacement $\vec{n}\to -\vec{n}$ and $s\to -s$\,. So the vertices $\vec{1}$ and
$\vec{n}\, \sigma_3$ are "even", whereas  $\sigma_3$ and $\vec{n}$ are "odd".

Using  relations (\ref{relat1}) we are able to present Eqs.~(\ref{LManalytic}) in the matrix form
 \be
&&\hat{\mathcal{T}}_0(\vec{n},q)=
\hat{\mathcal{T}}_0^\om(\vec{n},q) +
\langle\hat{\Gamma}_0(\vec{n},\vec{n\,}')\,\hat{\mathcal{G}}_4 (\vec{n'},q)\,
\hat{\mathcal{T}}_0(\vec{n'},q)\rangle_{\vec{n\,}'}\,,
\label{Leggett:T0}\\
&&\hat{\mathcal{T}}_0=\left[
\begin{array}{l}
\phantom{-}t_0^{(1)}\\ -t_0^{(2)}\\ \phantom{-}t_0^h\\ \phantom{-}t_0
\end{array}
\right],\,\,
\hat{\mathcal{T}}_0^\om= \left[\begin{array}{l}0\\0\\ t_0^{h\om}\\ t_0^{\om}\end{array}\right]
,\,\,
\hat{\Gamma}_0=\left[
\begin{array}{cccc}
\Gamma_0^\xi & 0 & 0 &0 \\
0 & \Gamma_0^\xi & 0 & 0 \\
0 & 0 & \Gamma_0^\om & 0 \\
0 & 0 & 0 & \Gamma_0^\om
\end{array}
\right], \nonumber\\ &&\hat{\mathcal{G}}_4 =\left[
\begin{array}{rrrr}
-G_{+}^h\cdot G_{-} & -F_+\cdot F_{-}  & G^h_{+}\cdot F_{-} & F_+\cdot G_{-}\\
 -F_+\cdot F_- & -G_{+}\cdot G_{-}^h & -F_+\cdot G_{-}^h & - G_{+}\cdot F_{-}\\
G^h_{+}\cdot F_{-} & -F_+\cdot G^h_{-} & G_{+}^h\cdot G^h_{-} & -F_+\cdot F_-\\
F_+\cdot G_{-} & -G_{+}\cdot F_- & -F_+\cdot F_- & G_{+}\cdot
G_{-}
\end{array}
\right]\,.
\label{Leggett:G}
 \ee
The matrix $\hat{\mathcal{G}}_4$ is precisely  the matrix used by Leggett in Eq.~(3) of
Ref.~\cite{L66}. The same matrix enters the equation for the vector vertices in (\ref{taut})
 \be
&&\hat{\vec{\mathcal{T}}}_1(\vec{n},q)=\hat{\vec{\mathcal{T}}}_1^\om(\vec{n},q) +
\langle\hat{\Gamma}_1(\vec{n},\vec{n\,}')\,\hat{\mathcal{G}}_4 (\vec{n'},q)\,
\hat{\vec{\mathcal{T}}}_1(\vec{n'},q)\rangle_{\vec{n\,}'}\,,
\label{Leggett:T1}\\
&&\hat{\vec{\mathcal{T}}}_1=\left[
\begin{array}{l}
\phantom{-}\vec{t\,}_1^{(1)}\\
-\vec{t\,}_1^{(2)}\\
-\vec{t\,}_1^h\\
\phantom{-}\vec{t\,}_1
\end{array}
\right],\,\,
\hat{\vec{\mathcal{T}}}_1^\om= \left[\begin{array}{l}0\\0\\ -\vec{t\,}^{h\om}\\
\phantom{-}\vec{t\,}^{\om}\end{array}\right]
,\,\,
\hat{\Gamma}_1=\left[
\begin{array}{cccc}
\Gamma_1^\xi & 0 & 0 &0 \\
0 & \Gamma_1^\xi & 0 & 0 \\
0 & 0 & \Gamma_1^\om & 0 \\
0 & 0 & 0 & \Gamma_1^\om
\end{array}
\right] .
\nonumber
 \ee
These equations elucidate the meaning of the
"arrow space" introduced by Leggett in Ref.~\cite{Leg65a} in application
to the in-medium vertices.
\begin{widetext}

To proceed further  we introduce even and odd
vertices. In our case the even vertices are
 \be
t^+={\txst\frac12} (t_0+t_0^h)\,,\quad
\vec{t}^{\,+}={\txst\frac12} (\vec{t}_1-\vec{t}_1^{\,h})\,,
\label{def:even}
\ee
and the odd ones are
\be
t^-={\txst\frac12} (t_0-t_0^h)\,,\quad
\vec{t}^{\,-}={\txst\frac12} (\vec{t}_1+\vec{t}_1^{\,h})\,.
\label{def:odd}
 \ee
With these definitions the bare vertices $\hat{1}$ and $\vec{n}\,\sigma_3$, considered by Leggett
in Ref.~\cite{L66}, contribute only to the even vertices, whereas $\vec{n}$ and $\sigma_3$ to the
odd vertices.

Equations for the even and odd amplitudes follow from Eqs.~(\ref{LManalytic}) if we take the
half-sum and the half-difference of Eqs.~(\ref{LMan:t0},\ref{LMan:th0}). Taking the
half-difference of Eqs.~(\ref{LMan:t10},\ref{LMan:t20}) and making use of Eq.~(\ref{t12relat}) we
obtain the equation for the anomalous vertex $\frac{1}{2}(t^{(2)}-t^{(1)})=t^{(2)}=\widetilde{t}$.
The systems (\ref{Leggett:T0}) and (\ref{Leggett:T1}) acquire then the same matrix form
 \be
&&\hat{\mathcal{T}}(\vec{n},q)=
\hat{\mathcal{T}}^\om(\vec{n},q) + \big\langle\hat{\Gamma}(\vec{n},
\vec{n\,})\hat{\mathcal{G}}_3 (\vec{n\,}',q)\,\hat{\mathcal{T}}(\vec{n\,}',q)\big\rangle_{\vec{n\,}'}
\label{Leggett:T}\\
&&\hat{\mathcal{T}}=\left[\begin{array}{l}-\widetilde{t}\\ -t^-\\ \phantom{-}t^+\end{array}\right]\,,
\quad
\hat{\mathcal{T}}^\om=\left[\begin{array}{l}\phantom{-}0\\ -t^{-\om}\\ \phantom{-}t^{+\om}\end{array}\right]
\,,\quad
\hat{\Gamma}=\left[\begin{array}{ccc}\Gamma^\xi & 0 & 0\\ 0 & \Gamma^\om & 0 \\ 0 & 0& \Gamma^\om\end{array}
\right]\,,
\nonumber\\
&&\hat{\mathcal{G}}_3 =\left[
\begin{array}{ccc}
-(G_{\scr +}\!\cdot\! G^h_{\scr -}+F_{+}\!\cdot\! F_{-}) &
G^h_{+}\!\cdot\! F_{-}-F_+\!\cdot\! G_{-} &
G^h_{+}\!\cdot\! F_{-} + F_{+}\!\cdot\! G_{-}\\
G^h_+\!\cdot\! F_- - F_+\!\cdot\! G_{-} &
{\txst\frac12}\big(G_{+}\!\cdot\! G_{-}+G^h_+\!\cdot\! G^h_-\big) + F_{+}\!\cdot\! F_{-} &
{\txst\frac12}\big(G^h_{+}\!\cdot\! G^h_{-} - G_+\!\cdot\! G_-\big)\\
G^h_+\!\cdot\! F_- + F_{+}\!\cdot\! G_{-} &
{\txst \frac12}\big(G^h_{+}\!\cdot\! G^h_{-} - G_+\!\cdot\! G_-\big) &
{\txst\frac12}\big(G_{+}\!\cdot\! G_{-}+G^h_+\!\cdot\! G^h_-\big)-F_{+}\!\cdot\! F_{-}
\end{array}
\right]\,
\label{gmatr}
 \ee
and we do not distinguish here the vertices with subscripts 0 and 1. In Eq.~(\ref{Leggett:T}) we
have to use $\Gamma^{\om(\xi)}=\Gamma^{\om(\xi)}_0$ for the vertices without $\sigma$ matrices
(subscript 0) and $\Gamma^{\om(\xi)}=\Gamma^{\om(\xi)}_1$ for the vertices with $\sigma$ matrices
(subscript 1). In Eq.~(\ref{gmatr}) we recognize the minor of the matrix $\hat{g}$ introduced by
Leggett, see Eq.~(12) in Ref.~\cite{L66}. Actually, Leggett in Ref.~\cite{L66} presented these
equations in a different form. To reproduce this form, following \cite{L66} we introduce new
quantities
 \be
\kappa(\vec{n\,},q) ={\txst\frac{1}{2}}\big(G_{+}^h\!\cdot\!
G^h_{-}+ G_{+}\!\cdot\! G_{-}\big)+ F_{+}\cdot F_{-}\,,\qquad
\lambda(\vec{n\,},q) = F_{+}\!\cdot\! F_{-}. \label{Legget:fun}
 \ee
All entries in the matrix (\ref{gmatr}) can be expressed through these two functions
 \begin{subequations}
\label{Legrel}
\be
G^h_{+}\cdot F_{-}- F_{+}\cdot
G_{-}&=&-\frac{\vec{v\,}\vec{q}}{\Delta}\, \lambda(\vec{n\,},q)\,,
\label{Legrel1}\\
G^h_{+}\cdot F_{-}+ F_{+}\cdot
G_{-}&=&\phantom{-}\frac{\om}{\Delta}\, \lambda(\vec{n\,},q)\,,
\label{Legrel2}\\
G_{+}\cdot G^h_{-}+ F_{+}\cdot F_{-}&=&A_0
+\frac{\om^2-(\vec{v\,}\vec{q\,})^2}{2\,\Delta^2}\,
\lambda(\vec{n\,},q),
\label{Legrel3}\\
{\txst\frac{1}{2}}\big(G_{+}^h\cdot G^h_{-}- G_{+}\cdot
G_{-}\big)&=&-\frac{\om}{\vec{v\,}\vec{q}}\,\kappa(\vec{n\,},q) .
\label{Legrel4}
\ee
 \end{subequations}
In~\cite{L66} Leggett emphasized that relations (\ref{Legrel1},\ref{Legrel2},\ref{Legrel3})) are
valid for any $\om$, $\vec{q}$ and $T$, whereas relation (\ref{Legrel4}) holds only in the limit
$v_\rmF\, |\vec{q\,}|,\,\om\ll\Delta$. In Appendix~\ref{app:loop} we re-derive these relations and
demonstrate that, actually, Eq. (\ref{Legrel4}) is valid for arbitrary $\om$  and $\vec{q}$.
Thereby, we prove that one may use these conditions in kinematic region of the PBF reactions.

With the help of Eq.~(\ref{Legrel}) we rewrite the system of equations
(\ref{Leggett:T}) as
 \be\label{gmatr1} &&\left[
\begin{array}{l}
-\widetilde{t}\\ -t^-\\ \phantom{-}t^+
\end{array}
\right] =\left[
\begin{array}{l}
\phantom{-}0\\ -t^{-\om}\\ \phantom{-}t^{+\om}
\end{array}
\right]+\left\langle\left[
\begin{array}{ccc}
\Gamma^\xi & 0 & 0\\ 0 & \Gamma^\om & 0 \\ 0 & 0& \Gamma^\om
\end{array}
\right]
\, \left[
\begin{array}{ccc}-A_0
+\frac{\om^2-(\vec{v\,}\vec{q})^2}{2\,\Delta^2}\,\lambda &
-\frac{\vec{v\,}\vec{q}}{\Delta}\,\lambda &
\frac{\om}{\Delta}\,\lambda
\\
-\frac{\vec{v\,}\vec{q}}{\Delta}\,\lambda &
\kappa &
 -\frac{\om}{\vec{v\,}\vec{q}}\,\kappa
\\
\frac{\om}{\Delta}\,\lambda  &
-\frac{\om}{\vec{v\,}\vec{q}}\,\kappa &
\kappa-2\,\lambda
\end{array}
\right]
\left[
\begin{array}{l}
-\widetilde{t}\\ -t^-\\ \phantom{-}t^+
\end{array}
\right]\right\rangle_{\vec{n\,}'}\, .\label{Leggett:kappa}
 \ee
Here we  recognize Eq.~(22) of Ref.~\cite{L66}, if we identify
 \be
t^{+\om}\rightarrow
\xi^{+}\,,\quad -t^{-\om}\rightarrow \xi^{-}\,,\quad
-\widetilde{t}\rightarrow \psi_1\,,\quad
-t^{-}\rightarrow \psi_2\,,\quad
t^+\rightarrow \psi_3\,.
\label{Legg:assign}
 \ee
Note that according to these assignments the quantities $\xi^\pm$ introduced in (22) by Leggett
ought to be defined as $\xi^{\pm}(\vec{n})=\frac{1}{2}[\xi (-\vec{p},-s )\pm \xi (\vec{p},s )]$\,.
Leggett defined $\xi^-$ with opposite sign. It has no influence on the correctness of his results,
since the quantities $\psi_{1,2,3}$ were not identified in~\cite{L66} with in-medium vertices and
the final expressions for the correlation functions, Eq.~(23) in Ref.~\cite{L66} depend quadratically
on $\xi^\pm$\,. Further, we shall name the system of equations (\ref{Leggett:kappa}), {\em the Leggett
equations.}

Now we formulate the current-current correlation function in terms
of the Leggett notations (using $t^{\pm}$ and $\widetilde{t}$
vertices). First, we express correlation functions
(\ref{chi:form}) in terms of vertices $t$, $t^h$ and
$\widetilde{t}$, cf. Eq.~(\ref{taut}),
 \be
&&\chi =\chi_0+\chi_1\,,
\nonumber\\
&&\chi_0 = \Big\langle\intop \rmd \Phi_T \,
t_0^\om(\vec{n},q)\, \Big(
G_+\, G_-\, t_0(\vec{n},q) - F_+\, F_- \,t_0^h(\vec{n},q) +(G_+\, F_- - F_+\, G_-)\,\widetilde{t}_0(\vec{n},q)
\Big)
\Big\rangle_{\vec{n}} \,,
\nonumber\\
&&\chi_1 = \Big\langle\intop \rmd \Phi_T \,
\vec{t}_1^{\,\om}(\vec{n},q)\, \Big(
G_+\, G_-\, \vec{t}_1(\vec{n},q)
+ F_+\, F_- \,\vec{t}_1^{\,h}(\vec{n},q)
+(G_+\, F_- - F_+\, G_-)\,\widetilde{\vec{t}}_1(\vec{n},q)
\Big)\Big\rangle_{\vec{n}}
\,.
\nonumber
 \ee
Further transformations we illustrate at hand of $\chi_0$. We do replacements $\epsilon_p\to
-\epsilon_p$ and $\epsilon_n\to -\epsilon_n$ ($\epsilon\to -\epsilon $ for $T=0$), which do not
change the integrals and the Matsubara sums, but induce replacements $F_{\pm}\to
F_{\mp}$\,,
$G_{\pm}\leftrightarrow G^h_{\mp}$ and $t\leftrightarrow t^h$. Then we add the resulting
expression to the initial one, divide by two and obtain
 \be
\chi_0 &=&
\frac12\Big\langle\intop \rmd \Phi_T \, t_0^\om(\vec{n},q)\, \Big(G_+\, G_-\, t_0(\vec{n},q) - F_+\, F_- \,t_0^h(\vec{n},q)
+(G_+\, F_- - F_+\, G_-)\,\widetilde{t}_0(\vec{n},q)\Big)\Big\rangle_{\vec{n}}
\nonumber\\
&+&\frac12\Big\langle\intop \rmd \Phi_T \, t_0^{h\om}(\vec{n},q)\, \Big(G^h_+\, G^h_-\, t^h_0(\vec{n},q) - F_+\, F_- \,t_0(\vec{n},q)
+(F_+\, G^h_- - G^h_+\, F_-)\,\widetilde{t}_0(\vec{n},q)\Big)\Big\rangle_{\vec{n}}\,.
\nonumber
\ee
The latter expression can be easily rewritten in terms of the even and odd vertices $t_0^\pm$ with
the result
 \be
\chi_0 &=& \Big\langle t^{+\om}_0\,
\Big[
(G^h_{+}\!\cdot\!F_{-}+F_{+}\!\cdot\!G_{-})\,(-\widetilde{t\,}_0)
+ {\txst\frac12}(G^h_{+}\!\cdot\!G^h_{-}-G_{+}\!\cdot\!G_{-})\,(-t_0^-)
+\big({\txst\frac12}(G_{+}\!\cdot\!G_{-}+G^h_{+}\!\cdot\!G^h_{-})-F_{+}\!\cdot\!F_{-}\big)\,
t_0^+ \Big] \Big\rangle_{\vec{n}}
\nonumber\\
&+&\Big\langle (-t^{-\om}_0)\, \Big[
(G^h_{+}\!\cdot\!F_{-}-F_{+}\!\cdot\!G_{-})\,(-\widetilde{t\,}_0)
+\big({\txst\frac12}(G_{+}\!\cdot\!G_{-}+G^h_{+}\!\cdot\!G^h_{-})+F_{+}\!\cdot\!F_{-}\big)\,(-t^-_0)
+ {\txst\frac12}(G^h_{+}\!\cdot\!G^h_{-}-G_{+}\!\cdot\!G_{-})\,t^+_0 \Big]
\Big\rangle_{\vec{n}}
\nonumber
 \ee
which in the Leggett's notations (\ref{Legget:fun}) turns into
\be
\chi_0&=& \chi_{0+} +\chi_{0-},
\nonumber\\
\chi_{0+}&=& \Big\langle t^{+\om}_0\,
\Big[
  \frac{\om}{\Delta}\lambda \,(-\widetilde{t\,}_0)
- \frac{\om}{\vec{v}\,\vec{q}}\,\kappa\, (-t^-_0)
+ (\kappa -2\lambda)\, t^+_0
\Big]
\Big\rangle_{\vec{n}},
\nonumber\\
\chi_{0-} &=& \Big\langle (-t^{-\om}_0)\,
\Big[
-\frac{\vec{v}\,\vec{q}}{\Delta}\,\lambda\,(-\widetilde{t\,}_0)
+ \kappa\, (-t^-_0)
- \frac{\om}{\vec{v}\,\vec{q}}\,\kappa\, t^+_0
\Big]
\Big\rangle_{\vec{n}}.
\label{chiLeggett}
 \ee
With replacements (\ref{Legg:assign}) we recover  Eqs.~(23a), and~(23b) of Ref.~\cite{L66}. Note that
the current-current correlator $\chi$ is given by the sum of the even and odd terms, whereas
Ref.~\cite{L66} presents two independent expressions, one $\propto \xi^{+}$ and another one $\propto
\xi^{-}$. For the vertices considered in Ref.~\cite{L66}  either the term $\propto
\xi^{+}$ is zero or that $\propto \xi^{-}$. But in general case the correct expression is
given by Eq.~(\ref{chiLeggett}). Similar equation holds also for the vertices with subscript
1:
 \be
\chi_1&=& \chi_{1+} +\chi_{1-},
\nonumber\\
\chi_{1+}&=&
\Big\langle
\vec{t\,\,}^{+\om}_1\,
\Big[
\frac{\om}{\Delta}\lambda\,(-\widetilde{\vec{t\,}}_1)
-\frac{\om}{\vec{v}\,\vec{q}}\,\kappa\, (-\vec{t\,}^-_1)
+ (\kappa-2\lambda)\, \vec{t\,}^+_1
\Big]
\Big\rangle_{\vec{n}},
\nonumber\\
\chi_{1-} &=& \Big\langle (-\vec{t\,\,}^{-\om}_1)\,
\Big[
-\frac{\vec{v}\,\vec{q}}{\Delta}\,\lambda\,(-\widetilde{\vec{t\,}}_1)
+ \kappa\, (-\vec{t\,}^-_1)
- \frac{\om}{\vec{v}\,\vec{q}}\,\kappa\, \vec{t\,}^+_1
\Big]
\Big\rangle_{\vec{n}}.
\label{chiLeggett1}
\ee
 \end{widetext}

Concluding this section we stress that both Larkin-Migdal and Leggett equations for
the vertices and the current-current correlators follow from the same set of equations, and, hence, the
results of the calculation of the emissivity in both approaches should be the same, provided
calculations are performed correctly. Now we will focus on the solution of these equations.

\section{Solution of Larkin-Migdal equations and  correlation functions}\label{sec:Solution}

\begin{widetext}

We apply the Larkin-Migdal equations (\ref{LMeq}) for the case of the weak-current vertices
(\ref{tauomh},\ref{full}). For the weak vector current vertices we use
Eqs.~(\ref{SwP:t0},\ref{SwP:tt0}) and for the weak axial-vector current vertices,
Eqs.~(\ref{SwP:t1},\ref{SwP:tt1}). Then we separate the parts proportional to the scalar $l_0$ and
to the vector $\vec{l}$ and obtain altogether 8 equations for vector and  axial-vector
current vertices. We cast these sets of equations in the following form~\cite{KV08},
 \begin{subequations}
\label{LMeqWeak}
\be
\tau_{a,0}(\vec{n},q)=\tau^{\om}_{a,0}(\vec{n},q)&+&
\big\langle\Gamma_a^{\om}(\vec{n},\vec{n'})\,
\big[
L(\vec{n'},q;\hat{\mathcal{P}}_{a,0})\, \tau_{a,0}(\vec{n'},q) +
M(\vec{n'},q)\,\widetilde{\tau}_{a,0}(\vec{n'},q)
\big]\big\rangle_{\vec{n}'}\,,
\label{LMEw1}\\
\widetilde{\tau}_{a,0}(\vec{n},q)=&-&
\big\langle\Gamma_a^{\xi}(\vec{n},\vec{n'})\,
\big[
(N(\vec{n'},q)+A_0)\, \widetilde{\tau}_{a,0}(\vec{n'},q) +
O(\vec{n'},q;\hat{\mathcal{P}}_{a,0})\,{\tau}_{a,0}(\vec{n'},q)
\big]\big\rangle_{\vec{n}'}\,,
\label{LMEw2}\\
\vec{\tau}_{a,1}(\vec{n},q)=\vec{\tau}^\om_{a,1}(\vec{n},q)&+&
\big\langle\Gamma_a^{\om}(\vec{n},\vec{n'})\,
\big[
L(\vec{n'},q;\hat{\mathcal{P}}_{a,1})\,\vec{\tau}_{a,1}(\vec{n'},q) +
M(\vec{n'},q)\,\vec{\widetilde{\tau}}_{a,1}(\vec{n'},q)
\big]\big\rangle_{\vec{n}'}\,,
\label{LMEw3}\\
\vec{\widetilde{\tau}}_{a,1}(\vec{n},q)=&-&
\big\langle\Gamma_a^{\xi}(\vec{n},\vec{n'})\,
\big[
(N(\vec{n'},q)+A_0)\, \vec{\widetilde{\tau}}_{a,1}(\vec{n'},q) +
O(\vec{n'},q;\hat{\mathcal{P}}_{a,1})\,\vec{\tau}_{a,1}(\vec{n'},q)
\big]\big\rangle_{\vec{n}'}\,.
\label{LMEw4}
 \ee
 \end{subequations}
To write one set of equations for both vector and axial-vector weak currents we introduced the
notation for the effective interaction $\Gamma_a^{\om,\xi}=\Gamma_0^{\om,\xi}$, if $a=V$, and
$\Gamma_a^{\om,\xi}=\Gamma_1^{\om,\xi}$, if $a=A$. Operators $\hat{\mathcal{P}}_{a,i}$ are defined
as follows
 \be
\hat{\mathcal{P}}_{a,i}=(-1)^i\, P_{a,i}\, \piproj\,,\quad i=0,1\,
 \ee
with  parameters
 \be
P_{V,0}=1\,,\quad P_{V,1}=-1\,,\quad P_{A,0}=-1\,,\quad
P_{A,1}=1\,,
 \ee
which are  eigenvalues of operators $\hat{\mathcal{P}}_{a,i}$,
when the latter are acting on the bare vertices
 \be
\hat{\mathcal{P}}_{a,0}\,\tau_{a,0}^\om=P_{a,0}\,\tau_{a,0}^\om\,,
\quad
\hat{\mathcal{P}}_{a,1}\,\vec{\tau}_{a,1}^\om=P_{a,1}\,\vec{\tau}_{a,1}^\om\,.
 \ee

To proceed  let us for simplicity assume that $\Gamma_a^{\om}$ and $\Gamma_a^{\xi}$ contain only
zero-th Legendre harmonics. Extension to higher harmonics will be done elsewhere~\cite{KVfuture}.
From (\ref{LMEw2}) we find\end{widetext}
 \be
\widetilde{\tau}_{a,0}(q) =-\eta^\xi_a\,
\frac{\big\langle O(\vec{n},q;P_{a,0})\big\rangle_{\vec{n}} }
{\big\langle N(\vec{n},q)\big\rangle_{\vec{n}}}\, {\tau}_{a,0}(q),
\label{widetilde} \ee
where
\be
\eta^\xi_a=\frac{\Gamma_a^\xi\,\langle N(\vec{n},q)\rangle_{\vec{n}} }
{1+\Gamma_a^\xi [A_0+ \langle N(\vec{n},q)\rangle_{\vec{n}}]}.
 \ee
For the channel $a$, for which the gap equation $1=-A_0\,\Gamma_{a}^{\xi} $ is valid, we obtain
$\eta_a^\xi=1$. For another channel, $\eta_a^\xi\neq 1$. For the s-wave pairing we consider in the
present paper, it holds $1=-A_0\,\Gamma_{V}^{\xi}$. In Ref.~\cite{KV08} we have put $\eta_a^\xi=1$
in both channels since owing to the identity $\big\langle
O(\vec{n'},q;-1)\big\rangle_{\vec{n}'}\equiv 0$ one gets $\widetilde{\tau}_{A,0}=0$ independently
on the assumed value of $\eta^\xi_A$, provided only zeroth harmonics of the interaction are
retained. Substituting Eq.~(\ref{widetilde}) in Eq.~(\ref{LMEw1}) we obtain
 \be
&&\tau_{a,0}(q) = \gamma_a(q;P_{a,0})\,\tau^\om_{a,0}\,,
\nonumber\\ &&\gamma_a^{-1}(q;P)=1 -
\Gamma^\om_a\,\langle\mathcal{L}_a(\vec{n},q;P)\rangle_{\vec{n}}\,,
\label{solLMeq}
 \ee
where we introduce the notation
\be
&&\mathcal{L}_a(\vec{n},q;P)= L(\vec{n},q;P) -
\eta_a^\xi\,
\frac{\langle  O(\vec{n},q;P)  \rangle_{\vec{n}}} {\langle N(\vec{n},q)\rangle_{\vec{n}}}\,
M(\vec{n},q)\, .
\label{curleL}
 \ee
Solving the second pair of the Larkin-Migdal equations (\ref{LMEw3},\ref{LMEw4}) we first note
that for the constant $\Gamma_a^{\om}$ and $\Gamma_a^{\xi}$ the angular averages on the right-hand
sides of  equations do not depend on $\vec{n}$. Therefore, the component of the bare vertex
proportional to $\vec{v}$ is not renormalized in medium. However, in view of the identity
 \be
\langle f(\vec{n},\vec{q\,})\, \vec{n}\rangle_{\vec{n}}= \langle
f(\vec{n},\vec{q\,})\, (\vec{n\,}\vec{q})\rangle_{\vec{n}}\,
\vec{q}/\vec{q\,}^2 \label{ident}
 \ee
valid for an arbitrary scalar function $f$ of $\vec{n}$ and $\vec{q}$, the full vertices gain a
component proportional to $\vec{q}$. Thus we decompose 3-vectors $\vec{\tau}_{a,1}(\vec{n},q)$ and
$\vec{\widetilde{\tau}}_{a,1}(\vec{n},q)$\, into the parts proportional to the  $\vec{n}$,
$\vec{n}_q =\vec{q}/|\vec{q}|$ vectors and introduce new scalar form factors
 \be
\vec{\tau}_{a,1}(\vec{n},q)={\tau}^{\om}_{a,1}\, \vec{n}+
{\tau}^{(q)}_{a,1}(q)\, \vec{n}_q ,
\quad
\vec{\widetilde{\tau}}_{a,1}(\vec{n},q)= {\widetilde{\tau}}^{(q)}_{a,1}(q)\, \vec{n}_q\,
\label{vecvert}
 \ee
with the bare vertex
 ${\tau}^{\om}_{a,1}=(\vec{n}\,\vec{\tau\,}^{\om}_{a,1})$\,.
Action of the operator $\hat{\mathcal{P}}_{a,1}$ on the vertices (\ref{vecvert}) is given by
\be
&&\hat{\mathcal{P}}_{a,1}\,
\vec{\tau}_{a,1}(\vec{n},q)=
P_{a,1}\,\tau_{a,1}^\om\,\vec{n}+ (-P_{a,1})\, \vec{n}_q\,
\tau_{a,1}^{(q)}(q),
\nonumber\\
&&\hat{\mathcal{P}}_{a,1}\,
\vec{\widetilde{\tau}}_{a,1}(\vec{n},q) = (-P_{a,1})\,\vec{n}_q\, \widetilde{\tau}_{a,1}^{(q)}(q)\, .
\nonumber
\ee
Then, from Eq.~(\ref{LMEw4}) we recover
 \be
\widetilde{\tau}_{a,1}^{(q)}= -\eta_a^\xi\,
\frac{\langle O(\vec{n}, q;-P_{a,1})\rangle_{\vec{n}}}{\langle N(\vec{n},q)\rangle_{\vec{n}}}\,
{\tau}_{a,1}^{(q)} -\eta_a^\xi\frac{\langle O(\vec{n}, q; P_{a,1}) (\vec{n\,}\vec{n}_q)\rangle_{\vec{n}}}
{\langle N(\vec{n}, q)\rangle_{\vec{n}}}\, \tau_{a,1}^\om\,.
\label{widetildevec}
 \ee
From Eq.~(\ref{LMEw3}), substituting there Eq.~(\ref{widetildevec}), we find
 \be
\tau^{(q)}_{a,1} &=& \gamma_a(q;-P_{a,1})\,\Gamma_a^\om\, \langle
\widetilde{\mathcal{L}}_a(\vec{n}, q; P_{a,1})
(\vec{n\,}\cdot\vec{n}_q) \rangle_{\vec{n}}\, \tau_{a,1}^\om\,.
\label{LME:sol:vec}
 \ee
Here we introduce the quantity
\be
\widetilde{\mathcal{L}}_a(\vec{n}, q; P)=
L(\vec{n}, q; P)-\eta^\xi_a\frac{\langle M(\vec{n},q)\rangle_{\vec{n}}}
{\langle N(\vec{n}, q)\rangle_{\vec{n}}}\,O(\vec{n}, q; P)
\label{curleLtilde}
 \ee
and use the identity
 \be\label{idL} \langle
\widetilde{\mathcal{L}}_a(\vec{n}, q; P_{a,1}) \rangle_{\vec{n}}=
\langle {\mathcal{L}}_a(\vec{n}, q; P_{a,1}) \rangle_{\vec{n}}\, ,
 \ee
which allows to use for the vector vertices the same function $\gamma_a$\, as for
the scalar vertex\,.

In terms of the loop-functions (\ref{depend}) the response functions (\ref{chi0},\ref{chi1})
can be expressed as
\be
&&\chi_{a,0}(\vec{n},q)= L(\vec{n},q;\hat{\mathcal{P}}_{a,0})\,\tau_{a,0}(\vec{n},q)+M(\vec{n},q)\,
\widetilde{\tau}_{a,0}(\vec{n},q),
\nonumber\\
&&\vec{\chi}_{a,1}(\vec{n},q)= L(\vec{n},q;\hat{\mathcal{P}}_{a,1})\,
\vec{\tau}_{a,1}(\vec{n},q)+M(\vec{n},q)\,
\vec{\widetilde{\tau}}_{a,1}(\vec{n},q).
\nonumber 
\label{chis} \ee
Using solutions (\ref{widetilde}) and (\ref{solLMeq}) for the
scalar vertices we find
\be
\chi_{a,0}(\vec{n},q) = \gamma_a(q;P_{a,0})\, \tau_{a,0}^\om \,\mathcal{L}
(\vec{n},q;P_{a,0})\,.
\label{chi0:sol}
 \ee
With the help of Eq. (\ref{vecvert}) we construct
\be
\vec{\chi}_{a,1}(\vec{n},q)=
L(\vec{n},q;P_{a,1})\vec{n}\,\tau_{a,1}^\om + L(\vec{n},q;-P_{a,1})\,\vec{n}_q\,
\tau^{(q)}_{a,1}(\vec{n},q) + M(\vec{n},q)\,\vec{n}_q\, \widetilde{\tau}^{(q)}_{a,1}(\vec{n},q)
\,.
\ee
Using solutions (\ref{widetildevec}) and (\ref{LME:sol:vec}) for the three-vector
vertices we obtain
\be
\vec{\chi}_{a,1}(\vec{n},q)
 &=&\Big(
L(\vec{n},q;P_{a,1})\,\vec{n}- M(\vec{n},q)\,\eta^\xi_a\,
\frac{\langle O(\vec{n}, q; P_{a,1}) (\vec{n\,}\vec{n}_q)\rangle_{\vec{n}}}
{\langle N(\vec{n}, q)\rangle_{\vec{n}}}\,\vec{n}_q\Big)\,\tau_{a,1}^\om \,
\nonumber\\
&+& \mathcal{L}_a(\vec{n},q;-P_{a,1})\,\gamma_a(q;-P_{a,1})\,\Gamma_a^\om\,
\langle \widetilde{\mathcal{L}}_a(\vec{n}, q; P_{a,1})(\vec{n\,}\vec{n}_q) \rangle_{\vec{n}}\,
\vec{n}_q\,\tau_{a,1}^\om\, ,
\label{chivec1:1}
 \ee
and, then, rewrite it as follows
\be
\vec{\chi}_{a,1}(\vec{n},q)&=&
\vec{\tau}^\om_{a,1}\,\gamma_a(q;-P_{a,1})\,{\mathcal{L}}_a(\vec{n},q;P_{a,1}) +
\delta\vec{\chi}_{a,1}(\vec{n},q)\,,
\label{chivec:sol}\\
\delta\vec{\chi}_{a,1}(\vec{n},q)&=&
\eta^\xi_a\, \frac{M(\vec{n},q)}{\langle N(\vec{n}, q)\rangle_{\vec{n}}}\,
\langle O(\vec{n'}, q; P_{a,1}) (\vec{n}-\vec{n\,}')
\rangle_{\vec{n'}}\, \tau_{a,1}^\om
\nonumber\\
&+& \gamma_a(q;-P_{a,1})\,\tau_{a,1}^\om\,\Gamma_a^\om\,\Big(
\mathcal{L}_a(\vec{n},q;-P_{a,1})\,
\langle \widetilde{\mathcal{L}}_a(\vec{n}, q; P_{a,1}) \vec{n\,} \rangle_{\vec{n}}
-\mathcal{L}_a(\vec{n},q;P_{a,1})\,\vec{n}\,
\langle \widetilde{\mathcal{L}}_a(\vec{n},q;-P_{a,1})\rangle_{\vec{n}}
\Big)\,.
\nonumber
 \ee
Please, pay attention to some misprints in Eqs.~(16,17) of Ref.~\cite{KV08}, which are now corrected  in
Eqs.~(\ref{widetildevec},\ref{LME:sol:vec},\ref{chi0:sol}) and
(\ref{chivec:sol}). The final expressions for the emissivity
in~\cite{KV08} remain unchanged.

In \cite{KV08} we have demonstrated the vector current conservation up to terms $\propto v_{\rm
F}^6$ for $T=0$. In the present work we are able to prove that the vector current is exactly
conserved for arbitrary temperatures. In Appendix~\ref{app:Ward} we prove relations
 \be
\langle \om\,
\chi_{V,0}-\vec{q}\,\vec{\chi}_{V,1}\rangle_{\vec{n}}=0 ,
\nonumber\\
\Im\langle (\vec{q}\,\vec{v}\,)\, (\om\, \chi_{V,0}-\vec{q}\,\vec{\chi}_{V,1})\rangle_{\vec{n}} =0\,.
\label{conserv}
 \ee
These relations ensure the transversality of the polarization tensor for the weak vector current
(\ref{PiV}), $\Pi_V^{\mu\nu}q_\nu=0=q_\mu\, \Pi_V^{\mu\nu}$\,.

The averages entering the quantities $K_V$ and $K_A$,
Eqs.~(\ref{kappaV:Ward}) and (\ref{kappaA}), which determine the neutrino emissivities (\ref{emiss}),
acquire now the following form
 \begin{subequations}
\label{aver}
 \be
\langle \chi_{a,0}(\vec{n},q)\rangle_{\vec{n}} &=& \gamma_a(q;P_{a,0})\, \tau_{a,0}^\om \,
\langle \mathcal{L}(\vec{n},q;P_{a,0})\rangle_{\vec{n}}\,,
\label{aver:chi0}\\
\langle \chi_{a,0}(\vec{n},q)\, (\vec{q\,}\vec{v}) \rangle_{\vec{n}} &=&
\gamma_a(q;P_{a,0})\, \tau_{a,0}^\om \,
\langle \mathcal{L}(\vec{n},q;P_{a,0})(\vec{q\,}\vec{v})\rangle_{\vec{n}}\,,
\label{aver:chi0qv}\\
\langle \vec{q\,} \vec{\chi}_{a,1}(\vec{n},q)\rangle_{\vec{n}}
&=& \gamma_a(q;-P_{a,1})\,
\langle
\widetilde{\mathcal{L}}_a(\vec{n},q;P_{a,1})\,(\vec{\tau\,}_{a,1}^\om\,\vec{q}\,)\rangle_{\vec{n}}\,.
\label{aver:chi1q}\\
\langle \vec{v\,} \vec{\chi}_{a,1}(\vec{n},q)\rangle_{\vec{n}} &=&
\langle L(\vec{n},q;P_{a,1})\, (\vec{v\,}\,\vec{\tau\,}_{a,1}^\om)\rangle_{\vec{n}} -
\langle M(\vec{n},q)\,(\vec{v\,}\vec{n}_q)\rangle_{\vec{n}}\,\eta^\xi_a\,
\frac{\langle O(\vec{n}, q; P_{a,1}) (\vec{\tau\,}_{a,1}^\om\,\vec{n}_q) \rangle_{\vec{n}}}
{\langle N(\vec{n}, q)\rangle_{\vec{n}}}
\nonumber\\
&+&
\gamma_a(q;-P_{a,1})\,\Gamma_a^\om\,
\langle \mathcal{L}_a(\vec{n},q;-P_{a,1})\,(\vec{v\,}\vec{n}_q)\rangle_{\vec{n}}\,
\langle \widetilde{\mathcal{L}}_a(\vec{n}, q; P_{a,1})(\vec{\tau\,}_{a,1}^\om\vec{n}_q) \rangle_{\vec{n}}
\,.
\label{aver:chi1v}
\ee
\end{subequations}

Let us now find the neutrino emissivity at the condition $v_{\rm F}\ll 1$, being valid in the
region of the $1S$ neutron pairing in neutron
stars~\cite{FRS76,VS87,SV87,MSTV90,SVSWW97,Minimal,KHY,YKL99,YLS99,V01,BGV04,GV05,PGW,Sedr07,KR,LP,SMS,KV08}.
First, we expand the averages of the $\mathcal{L}$ and $\widetilde{\mathcal{L}}$ functions
\begin{subequations}
\label{Lexp}
\be
\frac{1}{a^2\, \rho}\,\langle \mathcal{L}_a(\vec{n},q;+1)\rangle_{\vec{n}} &\approx&
\frac{\vec{q\,}^2 v_\rmF^2}{3\, \om^2}\,
\Big(1+\frac{3\vec{q\,}^2 v_\rmF^2}{5\, \om^2}\Big)\, g_T(0,0,0)
+\frac{v_\rmF^4\, q^4}{5\, \om^2}\, \Big(
 \frac{\partial g_T(0,0,0)}{\partial \om^2}
+\frac{\partial g_T(0,0,0)}{\partial (\vec{v\,}\vec{q})^2} \Big)
\nonumber\\
&-&\frac{4\,\vec{q}^{\,4} v_\rmF^4}{45\, \om^4}\,g_T(0,\om,0)
+ (\eta_a^\xi-1)\, \frac{\vec{q\,}^4 v_\rmF^4}{10}\,
\frac{\partial^2 g_T(0,0,0)}{\partial (\vec{v\,}\vec{q})^4}
\nonumber\\
&+& (\eta_a^\xi-1)\, \Big(1+\frac{\vec{q\,}^2 v_\rmF^2}{3\, \om^2}+\frac{\vec{q\,}^4 v_\rmF^4}{9\, \om^4}\Big)
\,g_T(0,\om,0)
\nonumber\\
&+&(\eta_a^\xi-1)\,\Big(1+\frac{3\vec{q\,}^2 v_\rmF^2}{5\, \om^2}\Big)\,
\frac{\vec{q\,}^2 v_\rmF^2}{3}\, \frac{\partial g_T(0,\om,0)}{\partial (\vec{v\,}\vec{q})^2}
+O(\vec{q\,}^6\, v_\rmF^6/\om^6)\,,
\label{Lexp:Llp}\ee
\be
\frac{1}{a^2\, \rho}\,\langle \mathcal{L}_a(\vec{n},q;-1)\rangle_{\vec{n}}&=&
\frac{1}{a^2\, \rho}\,\langle L(\vec{n},q;-1)\rangle_{\vec{n}}
\nonumber\\
&\approx&
\frac{\vec{q\,}^2 v_\rmF^2}{3\, \om^2}\,
\Big(1+\frac{3\,\vec{q\,}^2 v_\rmF^2}{5\, \om^2}\Big)\,
\big(g_T(0,0,0)-g_T(0,\om,0)\big)
\nonumber\\
&+&\frac{v_\rmF^4\, q^4}{5\, \om^2}\, \Big(
 \frac{\partial g_T(0,0,0)}{\partial \om^2}
+\frac{\partial g_T(0,0,0)}{\partial (\vec{v\,}\vec{q})^2} \Big)
\nonumber\\
&-&
\frac{\vec{q\,}^4 v_\rmF^4}{5\, \om^2}\,\frac{\partial g_T(0,\om,0)}{\partial (\vec{v\,}\vec{q})^2}
+O(\vec{q\,}^6\, v_\rmF^6/\om^6)\,,
\label{Lexp:Llm}\ee
\be
\frac{1}{a^2\, \rho}\,\langle \mathcal{L}_a(\vec{n},q;+1)\, (\vec{n\,}\vec{n\,}_q)\rangle_{\vec{n}} &=&
\frac{1}{a^2\, \rho}\,\langle \widetilde{\mathcal{L}}_a(\vec{n},q;-1)\, (\vec{n\,}\vec{n\,}_q)\rangle_{\vec{n}}
\nonumber\\
 &\approx&
\frac{|\vec{q}\,|\, v_\rmF}{3\, \om}\,
\Big(1+\frac{3\, \vec{q\,}^2\, v_\rmF^2}{5\, \om^2}\Big)\, g_T(0,0,0)
- \frac{4\, |\vec{q}\,|^3\, v_\rmF^3}{45\,\om^3}\, g_T(0,\om,0)
\nonumber\\
&+&\frac{v_\rmF^3\, q^3}{5\, \om}\, \Big(
 \frac{\partial g_T(0,0,0)}{\partial \om^2}
+\frac{\partial g_T(0,0,0)}{\partial (\vec{v\,}\vec{q})^2} \Big)
\nonumber\\
&+&
(\eta_a^\xi-1)\,\frac{|\vec{q}\,|\, v_\rmF}{3\, \om}\,
\Big(1+\frac{\vec{q\,}^2\, v_\rmF^2}{3\, \om^2}\Big)\, g_T(0,\om,0)
\nonumber\\
&+&
(\eta_a^\xi-1)\, \frac{|\vec{q}\,|^3\, v_\rmF^3}{5\, \om}
\,\frac{\partial g_T(0,\om,0)}{\partial (\vec{v\,}\vec{q})^2}
+O(\vec{q\,}^5\, v_\rmF^5/\om^5)\,,
\label{Lexp:Ll1p}\ee
\be
\frac{1}{a^2\, \rho}\,\langle \mathcal{L}_a(\vec{n},q;-1)\, (\vec{n\,}\vec{n\,}_q)\rangle_{\vec{n}} &=&
\frac{1}{a^2\, \rho}\,\langle \widetilde{\mathcal{L}}_a(\vec{n},q;+1)\, (\vec{n\,}\vec{n\,}_q)\rangle_{\vec{n}}
\nonumber\\
 &\approx&
\frac{|\vec{q}\,|\, v_\rmF}{3\, \om}\,
\Big(1+\frac{3\,\vec{q\,}^2\, v_\rmF^2}{5\, \om^2}\Big)\,\big(g_T(0,0,0)- g_T(0,\om,0)\big)
\nonumber\\
&+&\frac{v_\rmF^3\, q^3}{5\, \om}\, \Big(
 \frac{\partial g_T(0,0,0)}{\partial \om^2}
+\frac{\partial g_T(0,0,0)}{\partial (\vec{v\,}\vec{q})^2} \Big)
\nonumber\\
&-&\frac{|\vec{q}\,|^3\, v_\rmF^3}{5\, \om}
\,\frac{\partial g_T(0,\om,0)}{\partial (\vec{v\,}\vec{q})^2}
+O(\vec{q\,}^5\, v_\rmF^5/\om^5)\,,
\label{Lexp:Ll1m}
 \ee
 \end{subequations}
where $g_T (\vec{n},\om ,\vec{q})$ function is determined by Eqs.~(\ref{gT1}) of
Appendix \ref{app:loop} and
\be
g_T(0,0 ,0)=1-\intop_1^\infty \frac{\rmd y}{y^2\, \sqrt{y^2-1}}\,\frac{2}{e^{y\,\Delta/T}+1}\,.
\nonumber
\ee
From these results we immediately see that the correlation functions $\gamma_a$ differ from unity
only in the second order in $v_\rmF\,|\vec{q}\,|$, i.e.
\be
\gamma_a(q;P)\approx 1 +O(\Gamma_a^\om\,a^2\,\rho\, \vec{q\,}^2\, v_\rmF^2/\om^2)\,.
\label{gammaexp}
 \ee
From (\ref{Lexp:Llp}) and (\ref{Lexp:Llm}) we obtain that in the expression for the emissivity induced by
the vector currents (\ref{kappaV:Ward}) both scalar and vector components,
(\ref{aver:chi0}) and (\ref{aver:chi1v}), contribute at the order $v_\rmF^4$,
 \be
\Im\langle\chi_{V,0}(\vec{n},q) \rangle &\approx&
-\frac{4\,\vec{q\,}^4 v_\rmF^4}{45\, \om^4}\,e_V a\, \rho\,
\Im g_T(0,\om,0)\,,
\nonumber\\
\Im\langle \vec{v}\,\vec{\chi}_{V,1}(\vec{n},q) \rangle\, &\approx&
-\frac{2\,\vec{q\,}^2 v_\rmF^4}{9\,\om^2}\,e_V a\, \rho\,
\Im g_T(0,\om,0)\,.
\nonumber
 \ee
Working in the leading order in $v_\rmF$ we have to put $\gamma_V\to 1$ in view of
Eq.~(\ref{gammaexp}). Thus, generalization of the corresponding result of Ref.~\cite{KV08} to
arbitrary temperatures reduces in the leading order to the replacement $\Im g\to \Im g_T
(0,\om,0)$ with
 \be
\Im g_T(0,\om,0)=-\frac{2\,\pi\, \Delta^2\, \theta(\om-2\,\Delta)}{\om\, \sqrt{\om^2-4\, \Delta^2}}
\,\left(1-2\,n\big(\frac{\om}{2T}\big)\right).
\label{gt0}
 \ee
Finally, for the neutron PBF emissivity on the vector current we obtain (for one neutrino flavor)
 \be\label{eV}
\epsilon_{\nu\nu,V}^{\rm nPBF}&\simeq&\epsilon_{\nu\nu}^{(0n)}\,
g_V^2\,e_V^2\frac{4}{81}v_{\rmF,n}^4 \,, \quad g_V^2 e_V^2 =1.
 \ee
This is precisely  the result derived in Ref.~\cite{KV08}. The coefficient
$\frac{4}{81}v_{\rmF,n}^4$ distinguishes this result form that  previously obtained in
Ref.~\cite{FRS76} with the bare vertices,
 \be
\epsilon_{\nu\nu}^{(0n)}=
\frac{4 \rho\,G^2 \Delta^7_n}{15\, \pi^3}
I(\frac{\Delta}{T}),\,
I(z)=\!\intop^\infty_1\! \frac{\rmd y\, y^5}{\sqrt{y^2-1}}
\frac{1}{(e^{zy}+1)^2}\,.
\label{emiss0}
 \ee
Expression~(\ref{eV}) deviates only slightly from the corresponding result obtained in Ref.~\cite{LP}
with bare vertices.

Now let us turn to neutrino emissivity induced by the axial-vector current.
In the expansion $v_\rmF\ll 1$ the leading term contributing to the emissivity is of the order
$v_\rmF^2$. Keeping only the leading terms we cast Eq.~(\ref{kappaA}) as
\be
K_A \approx -g_A^2 e_A^2\,\rho\,v_\rmF^2\, \vec{q\,}^2\,
\big[1+(1-{\textstyle\frac23\, \frac{\vec{q}^{\,2}}{\om^2}})-
{\textstyle \frac23}\big]\, \Im g_T (0,\om,0).
\label{kappaA:2}
 \ee
The correlation factors $\gamma_a$ contribute at the sub-leading order $\sim v_\rmF^4$, therefore
we neglected these terms in approximate expression (\ref{kappaA:2}). We emphasize that the
last two cross terms in the squared brackets in Eq.~(\ref{kappaA}) cannot be eliminated.

Finally, for the neutron PBF emissivity induced by the axial-vector current
we obtain (for one neutrino flavor)
\be\label{eA}
\epsilon_{\nu\nu,A}^{\rm nPBF}&\simeq&
\left(1+\frac{11}{21}-\frac{2}{3}\right)\, g_A^{\,2}\,e_A^2
v_{\rmF,n}^2\,\epsilon_{\nu\nu}^{(0n)}\,.
 \ee
This again coincides with the result derived in \cite{KV08}. The resulting emissivity is the sum
of contributions (\ref{eV}) and (\ref{eA}). We stress that Eqs.~(\ref{eV}) and (\ref{eA}) are
approximate expressions obtained in the leading order in $v_{\rmF}$. General result looks more
cumbersome but it is easily recovered with the help of Eqs.~(\ref{aver}). The latter equations
are derived in the present paper at arbitrary temperature.

\section{Solution of  Leggett equations and correlation functions: Axial-vector current }\label{sec:Leggett}

In this section on example of  the weak axial-vector current we will show how to correctly apply
the Leggett formalism to calculate the current-current correlators. We  operate with equations
(\ref{Leggett:kappa}) and demonstrate explicitly how one should exploit the even and odd vertices
(\ref{def:even},\ref{def:odd}).

The bare vertex generated by the axial-vector current (\ref{AC:tauom}) gives rise only to the
contribution  to the $\vec{t}_1$ vertex. E.g., from (\ref{AC:t1}) we have
\be
\vec{t\,\,}^{\om}_1=- g_A\frac{e_A}{a}\, \big(\vec{v\,}\,l_0 - \vec{l\,}\big)\,,
\quad
\vec{t\,}_1=- g_A\, \big(\vec{\tau}_{A,1}(\vec{n},q)\,l_0 - \tau_{A,0}(\vec{n},q)\,\vec{l\,}\big)\,,
 \ee
and the anomalous vertex is given by
\be
\widetilde{\vec{t\,}}_1=- g_A\,
\big(\widetilde{\vec{\tau}}_{A,1}(\vec{n},q)\,l_0-
\widetilde{\tau}_{A,0}(\vec{n},q)\,\vec{l\,}\big)\,.
 \ee
In accordance with Eqs. (\ref{def:even},\ref{def:odd}) the even
and odd vertices are
\be
&&\vec{t\,\,}^{+\om}_1=-
g_A\,\frac{e_A}{a}\,\vec{v\,}\,l_0\,,
\quad
\vec{t\,\,}^+_1=- g_A\,\big(\vec{\tau\,}_{A,1}^-(\vec{n},q)\,l_0-\tau_{A,0}^-(\vec{n},q)\,\vec{l\,\,}\big)\,,
\nonumber\\
&&\vec{t\,\,}^{-\om}_1=\phantom{-} g_A\,\frac{e_A}{a}\,\vec{l\,}\,,\quad
\vec{t\,\,}^-_1=-g_A\,\big(\vec{\tau\,}_{A,1}^+(\vec{n},q)\,l_0-\tau_{A,0}^+(\vec{n},q)\,\vec{l\,\,}\big)\,;
\nonumber\\
&&\vec{\tau\,}_{A,1}^\pm(\vec{n},q)={\txst\frac12}\big(\vec{\tau\,}_{A,1}(\vec{n},q)\pm
\vec{\tau\,}_{A,1}(-\vec{n},q)\big)\,,
\nonumber\\
&&\tau_{A,0}^\pm(\vec{n},q)={\txst\frac12}\big(\tau_{A,0}(\vec{n},q)\pm
\tau_{A,0}(-\vec{n},q)\big)\,.
\label{Legeq:vert}
 \ee
Substituting these vertices in (\ref{Leggett:kappa}) and
separating the parts proportional to $l_0$ and $\vec{l}$ we arrive
at two sets of equations for the scalar and vector vertices. The
set  for the $\tau_{A,0}^\pm$ and $\widetilde{\tau}_{A,0}^\pm$
vertices is
 \be
&& \widetilde{\tau}_{A,0}=\Gamma_1^\xi\,\Big(
-A_0\,\widetilde{\tau}_{A,0}+
\big\langle{\txst\frac{\om^2-(\vec{v\,}'\vec{q})^2}{2\,\Delta^2}}\,\lambda(\vec{n\,}',q)
\big\rangle_{\vec{n\,}'}\,\widetilde{\tau}_{A,0} \,
-\big\langle{\txst\frac{\vec{v\,}'\vec{q}}{\Delta}}\,\lambda(\vec{n\,}',q)\big\rangle_{\vec{n\,}'}\,
\tau_{A,0}^+
-\big\langle{\txst\frac{\om}{\Delta}}\,\lambda(\vec{n\,}',q)\big\rangle_{\vec{n\,}'}\,
\tau_{A,0}^- \Big)\,,
\nonumber\\
&& \tau_{A,0}^+=\frac{e_A}{a}+\Gamma_1^\om\,\Big(
- \big\langle{\txst\frac{\vec{v\,}'\vec{q}}{\Delta}}\,\lambda(\vec{n\,}',q)\big\rangle_{\vec{n\,}'}\,
\widetilde{\tau}_{A,0}
+\langle\kappa(\vec{n\,}',q)\rangle_{\vec{n\,}'}\, \tau_{A,0}^+
+\big\langle{\txst\frac{\om}{\vec{v\,}'\vec{q}}}\,\kappa(\vec{n\,}',q)\big\rangle_{\vec{n\,}'}\,
\tau_{A,0}^-\Big)\,,
\nonumber\\
&& \tau_{A,0}^- =\Gamma_1^\om\,\Big(- \big\langle
{\txst\frac{\om}{\Delta}}\,\lambda(\vec{n\,}',q)\big\rangle_{\vec{n\,}'}\,\widetilde{\tau}_{A,0}
+\big\langle{\txst\frac{\om}{\vec{v\,}\vec{q}}}\,\kappa(\vec{n\,}',q)\big\rangle_{\vec{n\,}'}
\,\tau_{A,0}^+
+\langle\big(\kappa(\vec{n\,}',q)-2\lambda(\vec{n\,}',q)\big)
\rangle_{\vec{n\,}'}\,\tau_{A,0}^- \Big)\,.
 \label{Legeq:ta0-1}
 \ee
Here we  explicitly take into account that the vertices do not depend on $\vec{n}$, provided the
interaction constants $\Gamma_1^\om$ and $\Gamma_0^\xi$ contain only the zeroth Legendre harmonics.
Since  $\lambda$ and $\kappa$ are even functions of $\vec{v\,}\vec{q}$, these equations are
simplified as
 \be
&& \widetilde{\tau}_{A,0}=\Gamma_1^\xi\,\Big(
-A_0\,\widetilde{\tau}_{A,0}+
\big\langle{\txst\frac{\om^2-(\vec{v\,}'\vec{q})^2}{2\,\Delta^2}}\,\lambda(\vec{n\,}',q)
\big\rangle_{\vec{n\,}'}\,\widetilde{\tau}_{A,0}
-\big\langle{\txst\frac{\om}{\Delta}}\,\lambda(\vec{n\,}',q)\big\rangle_{\vec{n\,}'}\,
\tau_{A,0}^- \Big)\,,
 \nonumber\\
&& \tau_{A,0}^+=\frac{e_A}{a}+\Gamma_1^\om\, \langle\kappa(\vec{n\,}',q)\rangle_{\vec{n\,}'}\, \tau_{A,0}^+ ,
\nonumber\\
&& \tau_{A,0}^- =\Gamma_1^\om\,\Big(
-\big\langle{\txst\frac{\om}{\Delta}}\,\lambda(\vec{n\,}',q)\big\rangle_{\vec{n\,}'}\,\widetilde{\tau}_{A,0}
+\langle\big(\kappa(\vec{n\,}',q)-2\lambda(\vec{n\,}',q)\big)
\rangle_{\vec{n\,}'}\,\tau_{A,0}^- \Big)\,.
 \label{Legeq:ta0-2}
 \ee
The equation for $\tau_{A,0}^+$ decouples from  equations for $\tau_{A,0}^-$ and
$\widetilde{\tau}_{A,1}$. Then the solution is
 \be
{\tau\,}_{A,0}(\vec{n\,},q)=
{\tau\,}_{A,0}^+(\vec{n\,},q)+{\tau\,}_{A,0}^-(\vec{n\,},q)=\tau_{A,0}^+
={\txst\frac{e_A}{a}}\, \big[1-\Gamma_1^\om\,
\langle\kappa(\vec{n\,}',q)\rangle_{\vec{n\,}'}\big]^{-1} \,,\quad
\tau_{A,0}^-=\widetilde{\tau}_{A,0}=0\,.
\label{sol:tA0}
 \ee
Taking into account Eq.~(\ref{kappa-L}) we have $\langle\kappa(\vec{n\,}',q)\rangle_{\vec{n\,}'}=
\langle L(\vec{n\,}',q;-1)\rangle_{\vec{n\,}'}\equiv \langle
\mathcal{L}_A(\vec{n\,}',q;-1)\rangle_{\vec{n\,}'}$, since $\big\langle
O(\vec{n},q;-1)\big\rangle_{\vec{n}}\equiv 0$. Thus, we recover Eq.~(\ref{solLMeq}) with
$P_{A,0}=-1$\,. The anomalous vertex (\ref{widetilde}) vanishes for $P_{A,0}=-1$ in agreement with
Eq.~(\ref{Legeq:ta0-2}).

The second set of equations is for the vector vertices $\vec{\tau\,}_{A,1}^\pm$ and
$\widetilde{\vec{\tau}}_{A,1}$:
 \be
&& \widetilde{\vec{\tau}}_{A,1}(\vec{n\,},q)=\Gamma_1^\xi\,\Big(
-\langle A_0\,\widetilde{\vec{\tau}}_{A,1}(\vec{n\,}',q)\rangle_{\vec{n\,}'}
+ \big\langle{\txst\frac{\om^2-(\vec{v\,}'\vec{q})^2}{2\,\Delta^2}}\,\lambda(\vec{n\,}',q)
       \,\widetilde{\vec{\tau}}_{A,1}(\vec{n\,}',q)\big\rangle_{\vec{n\,}'}
-\big\langle{\txst\frac{\vec{v\,}'\vec{q}}{\Delta}}\,\lambda(\vec{n\,}',q)\,
     \vec{\tau\,}_{A,1}^{+}(\vec{n\,}',q)\big\rangle_{\vec{n\,}'}
\nonumber\\
&&\qquad\qquad
-\big\langle{\txst\frac{\om}{\Delta}}\,\lambda(\vec{n\,}',q)\, \vec{\tau\,}_{A,1}^-(\vec{n\,}',q)
    \big\rangle_{\vec{n\,}'} \Big)\,,
\nonumber\\
&&
\vec{\tau\,}_{A,1}^+(\vec{n\,},q)=\Gamma_1^\om\,
\Big(
\big\langle
{\txst\frac{\vec{v\,}'\vec{q}}{\Delta}}\,\lambda(\vec{n\,}',q)\, \widetilde{\vec{\tau}}_{A,1}(\vec{n\,}',q)
\big\rangle_{\vec{n\,}'}
+\langle\kappa(\vec{n\,}',q)\, \vec{\tau\,}_{A,1}^{+}(\vec{n\,}',q)\rangle_{\vec{n\,}'}
+\big\langle
{\txst\frac{\om}{\vec{v\,}'\vec{q}}}\,\kappa(\vec{n\,}',q)\,\vec{\tau\,}_{A,1}^-(\vec{n\,}',q)
 \big\rangle_{\vec{n\,}'}
\Big)\,,
\nonumber\\
&&
\vec{\tau\,}_{A,1}^-(\vec{n\,},q)=\frac{e_A}{a}\,\vec{v}
+\Gamma_1^\om\,\Big(
\big\langle
 {\txst\frac{\om}{\Delta}}\,\lambda(\vec{n\,}',q)\,\widetilde{\vec{\tau}}_{A,1}(\vec{n\,}',q)
\big\rangle_{\vec{n\,}'}
+\big\langle
  {\txst\frac{\om}{\vec{v\,}\vec{q}}}\,\kappa(\vec{n\,}',q)\, \vec{\tau\,}_{A,1}^{+}(\vec{n\,}',q)
\big\rangle_{\vec{n\,}'}
\nonumber\\
&&\qquad\qquad
+\langle\big(\kappa(\vec{n\,}',q)-2\lambda(\vec{n\,}',q)\big)\,
\vec{\tau\,}_{A,1}^-(\vec{n\,}',q) \rangle_{\vec{n\,}'} \Big)\,.
\nonumber
 \ee
The inspection of these equations reveals that the first and the third equations have trivial
solutions $\vec{\tau\,}_{A,1}^-= \frac{e_A}{a}\,\vec{v}$ and $\widetilde{\vec{\tau}}_{A,1}=0$\,.
Presenting the solution for the second equation  in the form $\vec{\tau\,}_{A,1}^+=\tau^+_q\,
\vec{n}_q$ we obtain
 \be
\tau^+_q &=&\Gamma_1^\om\,\big(
\langle\kappa(\vec{n\,}',q)\rangle_{\vec{n\,}'} \tau^+_q + \langle
\kappa(\vec{n\,}',q)\rangle_{\vec{n\,}'}
{\txst\frac{\om}{|\vec{q}|}\,\frac{e_A}{a}} \big)
\nonumber\\
&=&\Gamma_1^\om\,{\txst \frac{e_A}{a}\,\frac{\om}{|\vec{q\,}|}}\,
\langle \kappa(\vec{n\,}',q)\rangle_{\vec{n\,}'} \Big[1-
\Gamma_1^\om \,
\langle\kappa(\vec{n\,}',q)\rangle_{\vec{n\,}'}\Big]^{-1} \,.
\label{sol:tqp}
 \ee
In the last expression using first Eq.~(\ref{kappa-L}) and then Eqs.~(\ref{idL},\ref{eqLL}) we
obtain
 \be
\frac{\om}{|\vec{q\,}|}\, \langle \kappa(\vec{n\,}',q)\rangle_{\vec{n\,}'} =
\frac{\om}{|\vec{q\,}|}\, \langle \mathcal{L}_a(\vec{n\,}',q;-1)\rangle_{\vec{n\,}'} =
\frac{\om}{|\vec{q\,}|}\, \langle \widetilde{\mathcal{L}}_a(\vec{n\,}',q;-1)\rangle_{\vec{n\,}'} =
\langle \widetilde{\mathcal{L}}_a(\vec{n\,}',q;-1)\, (\vec{v\,}'\vec{n\,}_q)\rangle_{\vec{n\,}'}.
 \ee
Thus, we recover expressions (\ref{vecvert},\ref{LME:sol:vec}), which we have derived above within
the Larkin-Migdal formalism.

Now we  shortly dwell  on the axial-vector current-current correlators, as they follow from the
Leggett formalism. Substituting the vertices (\ref{Legeq:vert}) in Eqs.~(\ref{chiLeggett1})  for
the correlators and the corresponding polarization tensor terms we obtain
 \be
\chi_1^A&=& \chi_{1+}^A +\chi_{1-}^A ,
\nonumber\\
\chi_{1+}^A &=&
g_A^2\frac{e_A}{a}l_0^2\,\Big\langle \vec{v} \Big[
\frac{\om}{(\vec{v}\,\vec{q\,})}\,\kappa(\vec{n},q)\,\vec{\tau\,}_{A,1}^+(\vec{n},q)
+(\kappa(\vec{n},q)
-2\lambda(\vec{n},q))\,\vec{\tau\,}_{A,1}^-(\vec{n},q)
-\frac{\om}{\Delta}\lambda(\vec{n},q)
\,\widetilde{\vec{\tau}}_{A,1}(\vec{n},q) \Big]
\Big\rangle_{\vec{n}}
\nonumber\\
&-& g_A^2\frac{e_A}{a}l_0\,\vec{l}\,
\Big\langle \vec{v}
\Big[
 \frac{\om}{(\vec{v}\,\vec{q\,})}\,\kappa(\vec{n},q)\,\tau_{A,0}^+(\vec{n},q)
+ (\kappa(\vec{n},q)
- 2\lambda(\vec{n},q))\,\tau_{A,0}^-(\vec{n},q)
- \frac{\om}{\Delta}\lambda(\vec{n},q) \,\widetilde{\tau}_{A,0}(\vec{n},q)
\Big]
\Big\rangle_{\vec{n}}
\nonumber\\
&=& l_0^2\, \Pi_{00}^A(\om,\vec{q\,}) + l_0\, \vec{l}_i\, \Pi_{i0}^A (\om,\vec{q\,}),
\nonumber\\
\chi_{1-}^A &=&- g_A^2\,\frac{e_A}{a}\,\vec{l}\,l_0\,
\Big\langle\,
\Big[
\kappa(\vec{n},q)\,\vec{\tau\,}_{A,1}^+(\vec{n},q)
+\frac{\om}{(\vec{v}\,\vec{q\,})}\,\kappa(\vec{n},q)\,\vec{\tau\,}_{A,1}^-(\vec{n},q)
-\frac{(\vec{v}\,\vec{q\,})}{\Delta}\,\lambda(\vec{n},q)\,\widetilde{\vec{\tau}}_{A,1}(\vec{n},q)
\Big]
\Big\rangle_{\vec{n}}
\nonumber\\
&+&g_A^2\,\frac{e_A}{a}\,\vec{l\,}^2
\Big\langle\,
\Big[
\kappa(\vec{n},q)\,\tau_{A,0}^+(\vec{n},q)
+\frac{\om}{(\vec{v}\,\vec{q\,})}\,\kappa(\vec{n},q)\,\tau_{A,0}^-(\vec{n},q)
-\frac{(\vec{v}\,\vec{q\,})}{\Delta}\,\lambda(\vec{n},q)\,\widetilde{\tau}_{A,0}(\vec{n},q)
\Big]
\Big\rangle_{\vec{n}}
\nonumber\\
&=&
 l_0\, \vec{l}_i\, \Pi_{0i}^A (\om,\vec{q\,})
+\vec{l}_i\, \vec{l}_j\, \Pi_{ij}^A (\om,\vec{q\,}).
 \ee
The elements of the polarization tensor $\Pi_{A}$ derived here agree with those defined in Eq.~(\ref{PiA}) with
the correlators obtained within the Larkin-Migdal approach, (\ref{aver}), and the bare vertices
(\ref{AC:tauom})

We bring attention to the presence of the cross-terms $\Pi_{i0}^A$ and $\Pi_{0i}^A$, which were
introduced in Ref.~\cite{KV08}. These terms survive, even if we omit for simplicity all
correlation effects and put $\Gamma_1^\om=0$. In the latter case we get
 \be
\Pi_{0i}^A (\om,\vec{q\,}) &=& \Pi_{i0}^A (\om,\vec{q\,})=
-g_A^2\frac{e_A^2}{a^2}
 \Big\langle
\frac{\om \vec{v}_i}{(\vec{v}\,\vec{q\,})}\,\kappa(\vec{n},q)
 \Big\rangle_{\vec{n}}
\nonumber\\
&=&-g_A^2\frac{e_A^2}{a^2}\frac{\om\,\vec{q}_i}{\vec{q\,}^2}
\langle\kappa(\vec{n},q)\rangle_{\vec{n}}\,.
\label{cross-term:vac}
 \ee
Thus, we see that the cross-terms do not vanish, in contradiction   with the claim of
Ref.~\cite{LeinsonWrong}. The same cross-terms  are obtained from Eqs. (\ref{tauom}), (\ref{PiA}),
(\ref{chi0:sol}) and (\ref{chivec:sol}) within Larkin-Migdal formalism.

\section{Critical remarks to Ref.~\cite{LeinsonWrong}}\label{sec:Critic}

The results of Ref.~\cite{KV08} were  criticized by Leinson in Ref.~\cite{LeinsonWrong}, where the
emissivity of the PBF processes in the neutron superfluid with s-wave paring was recalculated in
the framework of the Leggett approach. The Fermi liquid amplitude of the $NN$ interaction
considered in Ref.~\cite{LeinsonWrong} included the zero and first Legendre harmonics in the
particle-hole channel. Comparison of the results of Ref.~\cite{LeinsonWrong} with our results
obtained in~Ref.~\cite{KV08} and here in Section~\ref{sec:Solution} reveals essential differences,
even if one keeps only zeroth harmonics in the particle-hole interaction and also  in the absence
of such interactions. Therefore we are forced to do some relevant comments.

(i) In Ref.~\cite{LeinsonWrong} the spatial component of the weak vector current  (the bare
vertex $\propto \vec{v}$) does not contain  the induced term $\propto \vec{q}/|\vec{q}\,|$. It can
be directly seen that such a term should appear in Eq.~(31) of Ref.~\cite{LeinsonWrong}, if we
substitute $\vec{\mathcal{T}}^-\propto \vec{v}$ in the third term of Eq.~(31) and use the relation
(\ref{ident}) above. In the same way the anomalous vertex $\vec{\tilde{\mathcal{T}}}$ gets the component
$\propto \vec{q}/|\vec{q}\,|$\,. The presence of these induced components of the in-medium
vertices is essential for establishing of the transversality of the polarization tensor and
proving of the vector current conservation. Without these terms the Ansatz (39)
in~\cite{LeinsonWrong} is invalid.

(ii) The temporal part of the axial-vector current (the bare vertex $\propto\vec{v}\vec{\sigma}$)
gets in Ref.~\cite{LeinsonWrong} the renormalization factor
\be
\frac{1}{1-\frac{g_0}{4} \langle
\kappa(\vec{n})-2\lambda(\vec{n})\rangle_{\vec{n}}}\approx
\frac{1}{1+\frac{g_0}{4} g_T(0,\om,0)
}
+O(g_0\,v_\rmF^2\, \vec{q\,}^2/\om^2)\,.
\label{Lein:wrong}
 \ee
The result (\ref{Lein:wrong}) is obviously incorrect, since the direction-independent interaction
(zeroth harmonics) cannot renormalize the vertex proportional to $\vec{n}$, the loops in
Fig.~\ref{fig:verteq} do not depend on $\vec{n}$\,.

From our calculations in Ref.~\cite{KV08} and here in Section~\ref{sec:Leggett} we see that only
the average $\langle \kappa(\vec{n})\rangle_{\vec{n}}\sim O(v_\rmF^2\, \vec{q\,}^2/\om^2)$ enters
the correlation factors in Eqs.~(\ref{sol:tA0}) and (\ref{sol:tqp}). We can argue differently. We
see that the first term on the right-hand side of Eq.~(\ref{Lein:wrong}) differs from unity even
in the limit $v_\rmF\to 0$. Assume now $\om \gg \Delta$. In this limit the correlation factor
should take the form as in the case of a normal Fermi liquid with the ordinary Lindhard's function
(taken now in the limit $\om \gg |\vec{q}|v_{\rm F}$). It is well known that the Lindhard function
is $\propto v_{\rm F}^2$ in this limit, see~\cite{VS87,MSTV90,M67} and Eq.~(\ref{gammaexp}) above.
This argument brought up in Ref.~\cite{KV08} was not perceived by the author of
Ref.~\cite{LeinsonWrong}.

In Ref.~\cite{LeinsonWrong} the author is surprised that in Ref.~\cite{KV08} we mentioned
importance of correlation effects but dropped them in the final expression for the emissivity. We
note that only in the expressions for the neutrino emissivities written up to the leading orders
in $v_\rmF$ (\ref{eV},\ref{kappaA:2},\ref{eA}) we can put $\gamma_a \simeq 1$.
This, obviously,
does not mean that $NN$ correlation factors can be always ignored.
General expressions for vertices in Ref.~\cite{KV08} contain correlation terms.
 Examples, where correlation terms give rise
to  important contributions,
 can be found in Refs.~\cite{MSTV90,KV08}.
 Also, in
Ref.~\cite{KV08} we mentioned a principal problem in applications of the BCS approximation to the
description of pairing in nuclear systems. In the BCS approximation one uses the same interactions
in particle-particle and particle-hole channels, whereas in the nuclear matter they can be
significantly different. Since the Migdal theorem (valid for the electron-phonon interaction) does
not hold in case of strongly interacting system, one should use the general Larkin-Migdal-Leggett
formalism. Arguing, we mentioned that in particle-particle channel the Landau-Migdal parameter is
necessary attractive $f_0^{\xi}<0$ (to provide s-pairing), but in the particle-hole channel one
has $f_0^{\om}>0$ at some densities, as follows from estimates of these parameters, see~\cite{M67}.

The origin of the mistake that have led in Ref.~\cite{LeinsonWrong} to Eq.~(\ref{Lein:wrong})  is
the wrong assignment of the bare vertex $\xi_\pm$ in Eq.~(91) there. The physical vertex cannot be
proportional to $|\vec{v}\,|$ which in the coordinate space would correspond to the operator
$\sqrt{\nabla^2}$ having bad analytical properties. Consequently, Eqs.~(96) and (97) in
Ref.~\cite{LeinsonWrong} are incorrect,  as being written for scalars but not for vector objects.
Following the original prescriptions of Ref.~\cite{Leg65a,L66}, for the temporal part of the
axial-vector current, $\vec{v}\,\vec{\sigma}$, we have to put $\xi_+=\vec{v}\,\vec{\sigma}$ and
$\xi_-=0$. Then the original Leggett equations (22) in Ref.~\cite{L66} yield the solution
 \be
\psi_1=0\,,\quad
\psi_2=\frac{\Gamma_1^\om \,\om\,\langle \kappa(\vec{n\,}')\rangle_{\vec{n\,}'}}
{1-\Gamma_1^\om\, \langle \kappa(\vec{n\,}')\rangle_{\vec{n\,}'}}\,
 \frac{ \vec{q}\,\vec{\sigma}}{\vec{q\,}^2}\,, \quad
\psi_3=\vec{v}\, \vec{\sigma}\,.
\nonumber
 \ee
For the spatial component of the axial-vector current, $\vec{\sigma}$, we have
$\vec{\xi}_+=0$ and $\vec{\xi}_-=\vec{\sigma}$ and the solutions are
\be
\vec{\psi}_1=\vec{\psi}_3=0\,,\quad
\vec{\psi}_2=\frac{\vec{\sigma}}
{1-\Gamma_1^\om \, \langle \kappa(\vec{n\,}') \rangle_{\vec{n\,}'}}
\,.
\nonumber
\ee

Thus, the main results of Ref.~\cite{LeinsonWrong} presented in Section~VII and in Fig.~1 are
false, since Eqs.~(118,119) and (125) are derived with the incorrect solutions of the Leggett
equations.

(iii) The author of Ref.~\cite{LeinsonWrong} claims the vanishing of the mixed term of the
polarization tensor $\Pi^{0i}$ and $\Pi^{i0}$, see Eq. ~(106) in~\cite{LeinsonWrong}. This
conclusion is drawn at hand of incorrect Eqs.~(98,99) in~\cite{LeinsonWrong}. In Eq.~(98) the left,
bare  vertex must be taken as $\vec{v}$ and should stand under the integral, then the first term
in the brackets gives non-vanishing result. In (99) the first term produces the finite value if
one uses the results of the correctly formulated and solved Leggett equations.

Further comments are in order:

Ref.~\cite{LeinsonWrong} pays attention to the essential temperature dependence of the imaginary
part of the retarded polarization function, which, as we have shown, is an artifact of the
incorrect
solution (\ref{Lein:wrong}). These corrections are small  in the limit $T\ll \Delta$. Also they
are small for $v_{\rmF}^2\ll1$ for arbitrary $T$. In general, the imaginary part of the full
retarded polarization function contains information not only on the processes of the one-nucleon
origin (i.e. the PBF processes) but also on all multi-nucleon processes, e.g. on the two-nucleon
$NN$- bremsstrahlung processes. It has been explained in Ref.~\cite{KV08} (see text before Eq.
(8)). In the limit $T\ll 2\Delta$ the phase space of different processes is well separated and
contributions of the PBF and the $NN$- bremsstrahlung processes are easily decoupled.  How to
perform generalization to arbitrary temperatures was mentioned in Ref.~\cite{KV08}, also rough
estimations of the dropped temperature dependent corrections are presented there, see discussion
after (13). The full expressions for the emissivities of the PBF processes obtained in the given
paper are obvious generalizations of the corresponding expressions of  Ref.~\cite{KV08}. For
$v_{\rm F}\ll 1$ the result of Ref.~\cite{KV08} proves to be correct in the leading order for all
temperatures $T<T_c$, provided for $I(z)$ one uses general expression (\ref{emiss0}) rather than
its low temperature limit.

Even if we artificially omit all $NN$ correlation effects, the result of Ref.~\cite{LeinsonWrong}
for the emissivity induced by the axial-vector current, Eq.~(121), disagrees with the result of
Ref.~\cite{KV08} (see Eq.~(34) in the latter work). The first difference is the factor $(m^*/m)^2$
in the first term in the round brackets of (121), which Ref.~\cite{LeinsonWrong} recovered in the
form   originally presented in Refs.~\cite{YKL99,KHY}. Ref.~\cite{KV08} argued that this term
should be replaced by unity. The factor $(m^*/m)^2$ appears, if one writes the temporal component
of the axial-vector current as for the bare current $\vec{\sigma}\vec{p}/m = \vec{v}\, m^*/m$,
where $\vec{v}$ is the velocity of the nucleon quasi-particles. If the same arguments are applied
to the spatial component of the vector current, there would be a problem with the conservation of
the vector current. Indeed, the Ward identity between the Green's functions of the quasiparticles
with energies $\epsilon_p=p^2/2\, m^*$ and the vertex is destroyed in this case (see a comment
after Eq.~(24) in \cite{KV08}). In general the central object of the study is the correlator of
the fully dressed, in-medium currents and, therefore, the bare quantities cannot enter the final
expressions. Appropriate excitations in Fermi liquids are quasiparticles obeying the Landau
kinetic equation, where the group velocity $v=(d\epsilon /dp)|_{p_{\rm F}}=p_{\rm F}/m^*$ enters
rather than the phase velocity $p_{\rm F}/m$. The proper Fermi liquid renormalization of the
vertices was worked out  in the seminal paper by Migdal~\cite{Migdal63}.

Ref.~\cite{LeinsonWrong} stresses that the second correction term, $11/21$, in Eq.~(121) was
obtained first in Ref.~\cite{FRS76}. Note that the correction found and used in previous works was
twice as small, and it was corrected in Ref.~\cite{KV08}. However the main difference between
Eq.~(121) in~\cite{LeinsonWrong}) and Eq.~(34) in~\cite{KV08} is the absence in Eq.~(121) of the
cross terms of the temporal and spatial components of the axial-vector current, which falls out in
Ref.~\cite{LeinsonWrong} because of the incorrect solution of the Leggett equations.

The factor $\frac{1}{4}$ accompanying the Landau-Migdal parameter $g_0$ in Eqs.~(100,101)
of Ref.~\cite{LeinsonWrong} with the origin in Eq.~(90) was taken as in  Ref.~\cite{L66}, where
$\vec{\sigma}$ stands for the operator of nucleon spin, whereas  in the amplitude and the bare
axial current in Ref.~\cite{LeinsonWrong}, see Eqs.~(1) and (89) there,
$\sigma_i$ enter as the Pauli matrices.

In spite of the special comment by Leggett that the relation (\ref{Legrel4}) is valid in the
limit $v |\vec{q}\,| , \om \ll \Delta$ only (see the comment before Eq.~(17) in Ref.~\cite{L66}),
Ref.~\cite{LeinsonWrong} uses this relation for $\om > 2\Delta$ without any explanation.

\section{Conclusion}\label{sec:Conclude}

In Ref.~\cite{LM63} Larkin and Migdal extended the Fermi liquid approach onto Fermi systems with
pairing. The equations for the full normal and anomalous vertices (the Larkin-Migdal equations)
have been derived. They considered a particular case of s-wave paring at zero temperature, aiming
at applications to atomic nuclei~\cite{Migdal59,ML64,Migdal64}. In Refs.~\cite{Leg65a,L66} Leggett
generalized this approach to the Fermi liquid at non-zero temperature and applied it to study the
low-frequency, low momenta collective excitations. In difference with Larkin and Migdal, Leggett
formulated equations in a matrix form for symmetric and antisymmetric vertices (the Leggett
equations). Explicit equations were formulated for vertices with the symmetry ($1$, $\sigma_3$,
$p_\mu$, $p_\mu \sigma_3$)\,.

The present analysis provides  necessary, although self-evident, extensions of the Larkin- Migdal
formalism~\cite{LM63} to arbitrary temperature and of the  Leggett formalism~\cite{Leg65a,L66} to
arbitrary frequencies and momenta and to the vector and axial-vector weak current symmetry.
Efficiency of the Larkin-Migdal and the Leggett approaches generalized  in such a way is
demonstrated on example of the calculation of the neutrino emissivity in the reactions of the
nucleon Cooper pair breaking and formation.  To be specific we considered s-state neutron pairing
and included only zero harmonics in the Fermi liquid interaction both in the particle-particle and
in the particle-hole channels. Compared to our previous work \cite{KV08}, where explicit expressions
were found in the low temperature limit $T\ll 2\Delta$, here we performed generalizations to
arbitrary temperatures. Since recently there were published  works, where presence of the formal
difference in the Larkin-Migdal and Leggett approaches have led to misleading conclusions, we
carefully analyzed both approaches demonstrating  explicitly  that they indeed lead to the very
same results.

First, from the diagrammatic equation for the current-current correlator and vertices we
reproduced the Larkin-Migdal and the Leggett equations. Within the Matsubara techniques for $T\neq
0$ we showed the correspondence between the vertices and the loop functions introduced in both
approaches. It turned out possible to cast all necessary loop functions in terms of the one master
function $g_T$. In passing we proved the validity of a relation between the loop functions at
arbitrary frequencies and momenta, which has been previously proven by Leggett in the low
frequency-momentum region.

Then we solved first the Larkin-Migdal and then the Leggett equations for the renormalized
vertices of the neutral weak currents  at arbitrary temperature and found the current-current
correlator for the neutrino-antineutrino pair  in the neutron superfluids, imaginary part of which
describes the neutrino-antineutrino pair production in reactions with the nucleon Cooper pair
breaking and formation. Also we proved the exact conservation of the vector current.

The results of recent publication~\cite{LeinsonWrong} based on the
Leggett approach are found to be invalid because of falsely
interpretation of the Leggett equations and their wrong solution.

\acknowledgements

We are grateful to D.~Blaschke and B.~Friman for the discussions. This work was partially
supported by COMPSTAR, an ESF Research Networking Programme, and by the German Research Foundation
DFG grants 436 RUSS 113/558/0-3 and WA 431/8-1.

\appendix
\section{The loop functions}
\label{app:loop}

At zero temperature the loop functions (\ref{depend}) were calculated in Ref.~\cite{VGL61} using
the Feynman method for the integral of the Green's function products
 \be
L(\vec{n},q; P) &=& a^2\, \rho\, \Big[\frac{\vec{q}\,
\vec{v}}{\om- \vec{q}\,\vec{v}}\,(1- g(z))-\frac{g(z)}{2}\,
(1+P)\Big]\,, \nonumber\\ M(\vec{n},q) &=& -a^2\,
\rho\,\frac{\om+\vec{q}\,\vec{v}}{2\, \Delta}\, g(z)\,,
\nonumber\\ N(\vec{n},q) &=& a^2\, \rho\,
\frac{\om^2-(\vec{q}\,\vec{v})^2}{4\, \Delta^2}\, g(z)\,,
\nonumber\\ O(\vec{n},q;P) &=& a^2\, \rho\, \Big[
\frac{\om+\vec{q}\,\vec{v}}{4\, \Delta}+
\frac{\om-\vec{q}\,\vec{v}}{4\, \Delta}\, P \Big]\, g(z)\,.
\label{loopsT0}
 \ee
 The universal function $g$ is  given by
 \be
g&=&2\intop\rmd \Phi_0\, \frac{F_{+}\,F_-}{a^2\, \rho}
=\intop
\frac{2\,\Delta^2\,\rmd\Phi_0}{[\epsilon_+^2-E_+^2]\,[\epsilon_-^2-E_-^2]}\,,
\nonumber
 \ee
where $\intop\rmd \Phi_0 =\intop\rmd \Phi_T$ for $T=0$, see Eq. (\ref{dotoper});
$\epsilon_\pm=\epsilon\pm{\textstyle \frac12}\om$ and $E_\pm=E_{p\pm q/2}$\,. Calculation of the
integral yields
 \be
g(z)&=&
 -\frac{{\rm arcsinh}\sqrt{z^2-1}}{z\,
\sqrt{z^2-1}}-\frac{i\, \pi\, \theta(z^2-1)}{2\, z\,
\sqrt{z^2-1}}\,, \label{gfunc} \\ z^2&=&\frac{\om^2-(\vec{q}\,
\vec{v})^2}{4\, \Delta^2}>1\,, \quad \vec{v}=v_\rmF \, \vec{n}.
\nonumber
 \ee
 At finite temperatures the Feynman method does not
work~\cite{W93} and the Matsubara techniques can be used instead:
\be
&&g_T(\vec{n},i\om_m,\vec{q}\,)=2\intop\rmd \Phi_T\,
\frac{F_{+}\,F_-}{a^2\, \rho} =2\,\Delta^2
\intop_{-\infty}^{+\infty}\rmd\epsilon_p\sum_{n=-\infty}^{+\infty}
\frac{T}{[(i\epsilon_n+i\om_m)^2-E_+^2]\,[(i\epsilon_n)^2-E_-^2]}
. \label{gT}
 \ee
 Here $\epsilon_n=(2\, n+1)\, \pi\, T$ and $\om_m=2\,
m\, \pi\, T$\,. Thus, we obtain
 \be
&&g_T(\vec{n},i\om_n,\vec{q}\,)= \Delta^2
\intop_{-\infty}^{+\infty}\rmd\epsilon_p \Bigg[ \frac{(E_+ -
E_-)}{E_+\, E_-}\, \frac{(n(E_-)-n(E_+))}{(i\om_m)^2-(E_+ -
E_-)^2} - \frac{(E_+ + E_-)}{E_+\, E_-}\, \frac{(1- n(E_-)-
n(E_+))}{(i\om_m)^2 - (E_+ + E_-)^2} \Bigg]\,, \label{gT1}
 \ee
with the fermion occupation function $n(x)=1/(\exp(x/T)+1)$\,. After the replacement $i\om_m\to
\om_+=\om+i\, 0$ we obtain the analytical continuation to the  retarded $g_T$ function in the $\om$
complex plain. Other Matsubara sums of products of the normal and anomalous Green's functions can
be obtained with the help of the following general relation
 \be
&&S^{ab}(\epsilon_p,\vec{n\,}, i\om_m\,\vec{q\,})=T\, \sum_{n}\,
\frac{X^a_{n,+}}{[(i\epsilon_n + i\om_m)^2-E_+^2]}\,
\frac{X^b_{n,-}}{[(i\epsilon_n)^2-E_-^2]}
\nonumber\\
&&=
\big(n (E_-)-n (E_+)\big)\,
\Big(\frac{Y^a_+\, Y^b_-}{i\om_m-E_+ +E_-}
    -\frac{ \overline{Y}^a_+\, \overline{Y}^b_-}{i\om_m + E_+ - E_-}
\Big)
\nonumber\\
&&+ \big(1-n (E_-)-n (E_+)\big)\,
\Big(\frac{\overline{Y}^a_+\, Y^b_-}{i\om_m+ E_+ + E_-}
     -\frac{ Y^a_+\,\overline{Y}^b_-}{i\om_m - E_+-E_-}
\Big),
\label{Smatr}
 \ee
where
 \be
&&X_{n,\pm}=\{\Delta,\,B_{n,\pm},\, \overline{B}_{n,\pm} \}
\,,\quad \nonumber\\
&&B_{n,+}=i\epsilon_n +i\om_m+\epsilon_{+}\,,\quad
  B_{n,-}=i\epsilon_n +\epsilon_{-}\,,
\nonumber\\
&&\overline{B}_{n,+}=-i\epsilon_n -i\om_m+\epsilon_{+}\,,\quad
  \overline{B}_{n,-}=-i\epsilon_n +\epsilon_{-}\,,
 \nonumber\\
&& Y_\pm=\{\Delta,A_\pm,\, \overline{A}_{\pm}\}/2\, E_\pm\,,\quad
\overline{Y}_\pm=\{\Delta,\overline{A}_\pm,\, A_{\pm}\}/2\, E_\pm\,,
 \nonumber\\
&& A_\pm=\epsilon_\pm + E_\pm\,,\quad
\overline{A}_\pm = \epsilon_\pm - E_\pm\,,\quad
\epsilon_\pm=\epsilon_{\vec{p}\pm\vec{q}/2}\,.
 \nonumber
 \ee
Convolutions of the Green's functions can be expressed through the corresponding elements of the
matrix $S^{ab}$ integrated of $\rmd \epsilon_p$, e.g.,
 $$
 (F_+\!\cdot\! G^h_-)(\vec{n},\om,\vec{q}\,)=
a^2\, \rho \intop_{-\infty}^{+\infty}\rmd \epsilon_p S^{13}(\epsilon_p,\vec{n}, \om,\vec{q}\,) \,.
 $$
The matrix (\ref{Smatr}) has the following properties
 \be
S^{ab}(\epsilon_p,\vec{n},\om,\vec{q})=
S^{ba}(-\epsilon_p,\vec{n},-\om,-\vec{q})\,,
\label{Sexchange}
\ee
and since $Y_\pm^{(2)}=\overline{Y}_\pm^{(3)}$
 \be
S^{33}(\epsilon_p,\vec{n},\om,\vec{q}\,)=
S^{22}(\epsilon_p,\vec{n},-\om,\vec{q}\,)\,.
\label{Shexchange}
 \ee
Relation (\ref{Sexchange}) allows  to interchange the order of Green's function in the
convolutions with the simultaneous change of $\om\to-\om$ and $\vec{q}\to -\vec{q}$, e.g.,
 \be
&&(F_+\!\cdot\! G^h_-)(\vec{n},\om,\vec{q}\,)=(G^h_+\!\cdot\! F_-)(\vec{n},-\om,-\vec{q}\,)\,,
\nonumber\\
&&(G_+\!\cdot\! G^h_-)(\vec{n},\om,\vec{q}\,)=(G^h_+\!\cdot\! G_-)(\vec{n},-\om,-\vec{q}\,)\,,
\nonumber\\
&&(F_+\!\cdot\! G_-)(\vec{n},\om,\vec{q}\,)=(G_+\!\cdot\! F_-)(\vec{n},-\om,-\vec{q}\,)\,.
 \label{GFexch}
 \ee
 From
(\ref{Shexchange}) we find
\be
(G^h_+\!\cdot\!G^h_-)(\vec{n},\om, \vec{q}\,)=(G_+\!\cdot\!G_-)(\vec{n},-\om, \vec{q}\,)\,.
\label{GhGhrelat}
 \ee
Now we are able to consider the Green's function products entering
the matrix (\ref{Leggett:G}). We find, see \cite{KR,SMS},
\be
(G_{+}\!\cdot\!G_{-})(\vec{n},\om, \vec{q\,})&=&a^2\, \rho
\intop_{-\infty}^{+\infty}\rmd \epsilon_p\, \Big[
\frac{n(E_-)-n (E_+)}{2\,E_+\,E_-}\,
\frac{(\epsilon_{+}\,E_- +E_+\,\epsilon_{-})\,\om+ (\epsilon_+\,\epsilon_- + E_+\,E_-)(E_+ - E_-)}
{\om^2 - (E_+ - E_-)^2}
\nonumber\\
&&\qquad\quad
+\frac{1- n (E_-)- n (E_+)}{2\,E_+\,E_-}\,
\frac{
(\epsilon_+\, E_- - E_+\,\epsilon_-)\,\om -(\epsilon_+\,\epsilon_- - E_+\, E_-)\,(E_+ + E_-)
}{\om^2 - (E_+ + E_-)^2}
\Big]\,,
\label{Mats:GG}
 \ee
 \be
(G_{+}\!\cdot\!F_{-})(\vec{n},\om, \vec{q\,})&=&
-a^2\, \rho\,\Delta\intop_{-\infty}^{+\infty}\rmd \epsilon_p
\Big[
\frac{n(E_-)-n(E_+)}{2\,E_+\, E_-}
\frac{\om\,E_+ + \epsilon_+\,(E_+-E_-)}{\om^2-(E_+ - E_-)^2}
\nonumber\\
&&\qquad\quad+\frac{1-n(E_-)-n(E_+)}{2\,E_+\, E_-}\,
\frac{-\om\,E_+ - \epsilon_+\,(E_++E_-)}{\om^2 - (E_+ + E_-)^2}
\Big]\,,
\label{Mats:GF}
 \ee
 \be
(F_{+}\!\cdot\!G_{-})(\vec{n},\om, \vec{q\,})&=&
-a^2\,\rho\,\Delta\intop_{-\infty}^{+\infty}\rmd \epsilon_p
\Big[
\frac{n(E_-)-n(E_+)}{2\,E_+\, E_-}\,
\frac{\om\,E_- + \epsilon_-\,(E_+-E_-)}{\om^2-(E_+ - E_-)^2}
\nonumber\\
&&\qquad\quad
+\frac{1-n(E_-)-n(E_+)}{2\,E_+\, E_-}\,
\frac{\om\,E_- - \epsilon_-\,(E_++E_-)}{\om^2 - (E_+ + E_-)^2}
\Big]\,.
 \label{Mats:FG}
 \ee
Making the replacement $\epsilon_p\to -\epsilon_p$  in the integral (\ref{Mats:GF}), which induces
the changes $\epsilon_\pm\to -\epsilon_\mp$ and $E_\pm\to E_\mp$, we obtain exactly the same
integral as in Eq.~(\ref{Mats:FG}) but with  opposite sign. Thus, we prove that
 \be
 (G_{+}\!\cdot\!F_{-})(\vec{n},\om, \vec{q\,})&=&-(F_{+}\!\cdot\!G_{-})(\vec{n},\om, \vec{q\,})\,.
\label{relation1}
  \ee
For the $(G_{+}\!\cdot\!G^h_{-})$ loop we have
\be
(G_{+}\!\cdot\!G^h_{-})(\vec{n},\om, \vec{q\,})&=&
-a^2\,\rho\intop_{-\infty}^{+\infty}\rmd \epsilon_p \Big[
\frac{n(E_-)-n(E_+)}{2\,E_+\, E_-}\,
\frac{(\epsilon_-\, E_+-\epsilon_+\, E_-)\,\om+ (\epsilon_+\,\epsilon_--E_-\, E_+)\, (E_- -E_+) }
     {\om^2-(E_+ - E_-)^2}
\nonumber\\
&+&\frac{1-n(E_-)-n(E_+)}{2\,E_+\, E_-}\,
\frac{-(\epsilon_-\, E_++\epsilon_+\, E_-)\,\om + (\epsilon_-\, \epsilon_+ + E_+\, E_-)\, (E_++E_-)}
     {\om^2 - (E_+ + E_-)^2} \Big]\,.
\label{Mats:GGh}
 \ee
It is easy to verify that this integral does not change under the replacement $\vec{q}\to
-\vec{q}$, since then $\epsilon_\pm\to \epsilon_\mp$ and $E_\pm\to E_\mp$, and therefore
$(G_{+}\cdot G^h_{-})(\vec{n},\om, \vec{q\,})=(G_{+} \cdot G^h_{-})(\vec{n},\om, -\vec{q\,})$\,.
One can see that the replacement $\epsilon_p\to -\epsilon_p$ in the integral is equivalent to the
replacement $\om\to -\om$. Hence we prove the relation $(G_{+} \cdot G^h_{-})(\vec{n},-\om,
\vec{q\,})=(G_{+}\cdot G^h_{-})(\vec{n},\om, \vec{q\,})$\,. Combining the last two equations with
(\ref{GFexch}) we prove
 \be
(G_{+} \cdot G^h_{-})(\vec{n},\om, \vec{q\,})=
(G_{+} \cdot G^h_{-})(\vec{n},-\om, -\vec{q\,})=
(G^h_{+} \cdot G_{-})(\vec{n},\om, \vec{q\,}) .
\label{relation2}
  \ee
Now we state the results for $(F_{+}\!\cdot\!G^h_{-})$ and
$(G^h_{+}\!\cdot\,F_{-})$ products
\be
(F_{+}\!\cdot\!G^h_{-})(\vec{n},\om, \vec{q\,})&=&
-a^2\, \rho\, \Delta \intop_{-\infty}^{+\infty}\rmd \epsilon_p
\Big[
\frac{n(E_-)-n(E_+)}{2\,E_+\, E_-}\,
\frac{-E_-\,\om+\epsilon_-\, (E_+-E_-)}{\om^2-(E_+ - E_-)^2}
\nonumber\\
&+&\frac{1-n(E_-)-n(E_+)}{2\,E_+\, E_-}
\frac{-E_-\, \om - \epsilon_-\, (E_++E_-)}{\om^2 + (E_+ + E_-)^2}
\Big]\,,
\label{Mats:FGh}
\ee
 \be
(G^h_{+}\!\cdot\!F_{-})(\vec{n},\om, \vec{q\,})&=&
-a^2\, \rho\, \Delta \intop_{-\infty}^{+\infty}\rmd \epsilon_p \Big[
\frac{n(E_-)-n(E_+)}{2\,E_+\, E_-}\,
\frac{-E_+\,\om+\epsilon_+\, (E_+-E_-)}{\om^2-(E_+ - E_-)^2}
\nonumber\\
&+&\frac{1-n(E_-)-n(E_+)}{2\,E_+\, E_-}
\frac{E_+\, \om -\epsilon_+\, (E_++E_-)}{\om^2 + (E_+ + E_-)^2}
 \Big]\,.
\label{Mats:GhF}
 \ee
Making use of the replacement $\epsilon_p\to -\epsilon_p$ as in Eqs.~(\ref{Mats:GF},\ref{Mats:FG})
we are able to  prove that
 \be
(F_{+}\!\cdot\!G^h_{-})(\vec{n},\om, \vec{q\,})= -(G^h_{+}\!\cdot\!F_{-})(\vec{n},\om, \vec{q\,}).
\label{relation3}
 \ee
Eqs. (\ref{relation1}),  (\ref{relation2}) and (\ref{relation3}) are used in the paper body, see
Eq. (\ref{relat1}).

Now we turn to the derivation of relations (\ref{Legrel}). Consider first (\ref{Legrel1}). From
(\ref{Mats:FG}) and (\ref{Mats:GhF}) we have
 \be
(G^h_{+}\!\cdot\!F_{-})-(F_{+}\!\cdot\!G_{-})&=&
-a^2\, \rho\, \Delta \intop_{-\infty}^{+\infty}\rmd \epsilon_p \Big[
\frac{n(E_-)-n(E_+)}{2\,E_+\, E_-}\,
\frac{-(E_++E_-)\,\om+(\epsilon_+-\epsilon_+)\, (E_+-E_-)}{\om^2-(E_+ - E_-)^2}
 \nonumber\\
&+&\frac{1-n(E_-)-n(E_+)}{2\,E_+\, E_-}
\frac{(E_+-E_-)\, \om - (\epsilon_+-\epsilon_-)\, (E_++E_-)}{\om^2 + (E_+ + E_-)^2}
\Big]\,.
\nonumber
 \ee
The terms with $\om$ in the numerator vanish exactly since they are  antisymmetric with respect to
the replacement $\epsilon_p\to -\epsilon_p$. In the remaining terms using
$\epsilon_+-\epsilon_-=\vec{v\,}\vec{q}$ and comparing with (\ref{gT1}) and (\ref{gT}) we find
 \be
(G^h_{+}\!\cdot\!F_{-})-(F_{+}\!\cdot\!G_{-})=
-a^2\, \rho\, \frac{\vec{v\,}\vec{q}}{2\,\Delta}\, g_T(\vec{n\,},\om,\vec{q\,})=
-\frac{\vec{v\,}\vec{q}}{\Delta}\,(F_{+}\!\cdot\!F_{-})\,.
\nonumber
 \ee
Similarly, using the symmetry properties of the integrand under the change $\epsilon_p\to
-\epsilon_p$ we verify that in the sum $(G^h_{+}\cdot F_{-})+(F_{+} \cdot G_{-})$ only the term
with $\om$ will survive in the numerator
 \be
(G^h_{+}\!\cdot\!F_{-})+(F_{+}\!\cdot\!G_{-})&=&
-a^2\, \rho\, \Delta \intop_{-\infty}^{+\infty}\rmd \epsilon_p
\Big[
\frac{n(E_-)-n(E_+)}{2\,E_+\, E_-}\,
\frac{-(E_+-E_-)\,\om}{\om^2-(E_+ - E_-)^2}
\nonumber\\
&+&\frac{1-n(E_-)-n(E_+)}{2\,E_+\, E_-}
\frac{(E_++E_-)\, \om}{\om^2 + (E_+ + E_-)^2}
\Big]
\nonumber\\
&=&a^2\, \rho\, \frac{\om}{2\,\Delta}\,g_T(\vec{n\,},\om,\vec{q\,})
=\frac{\om}{\Delta}\, (F_{+}\!\cdot\!F_{-})\,.
\nonumber
 \ee
Thus, we recovered Eq.~(\ref{Legrel2}).

Before we consider  relations (\ref{Legrel3}) and (\ref{Legrel4}) let us simplify the numerators
in (\ref{Mats:GG}) and (\ref{Mats:GGh}). We use
 \be
\epsilon_\pm=\epsilon_p\pm{\txst \frac12}
\,(\vec{v\,}\vec{q})\,,\quad {\txst
\frac12}(E_+^2+E^2_-)=\epsilon_p^2+ {\txst \frac14}\,
(\vec{v\,}\vec{q})^2+\Delta^2\,, \quad{\txst
\frac12}(E_+^2-E^2_-)=\epsilon_p\,(\vec{v\,}\vec{q})\,,
\label{smplf}
 \ee
 and obtain
 \be
&&(\epsilon_{+}\,E_- \pm E_+\,\epsilon_{-})\,\om -
(\epsilon_+\,\epsilon_- \pm E_+\,E_-)\,(E_- \mp E_+) \nonumber\\
&&\qquad\qquad=(E_- \mp E_+)\Big( \Delta^2 +\big((\vec{v\,
}\vec{q\,})^2-(E_-\pm E_+)^2\big)\frac{\om
+(\vec{v\,}\vec{q\,})}{2\, (\vec{v\, }\vec{q})}\Big) .
\label{numerators}
 \ee
From Eq.~(\ref{GhGhrelat}) follows that $(G^h_+\cdot
G^h_-)-(G_+\cdot G_-)$ is an odd function of $\om$ and  contains
only the terms from (\ref{numerators}) linear in $\om$.
Oppositely, $(G^h_+\cdot G^h_-)+(G_+\cdot G_-)$ is an  even
function of $\om$ and contains the terms  from (\ref{numerators})
independent of $\om$. Hence we can write
 \be
\frac12(G^h_{+}\cdot G^h_{-}-G_{+}\cdot G_{-})
&=&-\frac{\om}{2\,(\vec{v\,}\vec{q\,})}\,
a^2\, \rho\intop_{-\infty}^{+\infty}\rmd \epsilon_p\, \Big[
\frac{n (E_-)-n (E_+)}{2\,E_+\,E_-}\, (E_- - E_+)
\frac{(\vec{v\,}\vec{q\,})^2-(E_+ +E_-)^2}{\om^2 - (E_+ - E_-)^2}
 \nonumber\\
&+&\frac{1- n (E_-)- n (E_+)}{2\,E_+\,E_-}\, (E_+ + E_-)\,
\frac{(\vec{v\,}\vec{q\,})^2-(E_+ - E_-)^2}{\om^2 - (E_+ + E_-)^2} \Big]\, ,
\label{GhGh-GG}
 \ee
and
 \be
\frac12(G^h_{+}\cdot G^h_{-}+G_{+}\cdot G_{-})+(F_+\cdot F_-)
&=&a^2\, \rho\,\frac12\intop_{-\infty}^{+\infty} \rmd \epsilon_p\, \Big[
\frac{n (E_-)-n (E_+)}{2\,E_+\,E_-}\,(E_- - E_+)
\frac{(\vec{v\,}\vec{q\,})^2-(E_+ +E_-)^2}{\om^2 - (E_+ - E_-)^2}
\nonumber\\
&+&\frac{1- n (E_-)- n (E_+)}{2\,E_+\,E_-}\, (E_+ + E_-)\,
\frac{(\vec{v\,}\vec{q\,})^2-(E_+ - E_-)^2}{\om^2 - (E_+ + E_-)^2}
\Big]\,.
\label{GhGh+GG}
\ee
Comparing Eqs.~(\ref{GhGh-GG}) and (\ref{GhGh+GG}) we conclude that
\be
{\txst\frac12}\,(G^h_{+}\cdot G^h_{-}-G_{+}\cdot G_{-})=
-\frac{\om}{(\vec{v\,}\vec{q\,})}\,\Big[{\txst\frac12}\,(G^h_{+}\cdot
G^h_{-}+G_{+}\cdot G_{-})+(F_+\cdot F_-)\Big]\,,
 \ee
thus, recovering Eq. (\ref{Legrel4}).

We turn now to the derivation of the relation (\ref{Legrel3}). First, we note that in
(\ref{Mats:GGh}) the terms with $\om$ in the numerator do not contribute because of their symmetry
properties with respect to the change of the sign of $\epsilon_p$. In the other terms in the numerator
we can use
 \be
\epsilon_+\,\epsilon_-\pm E_+\, E_- ={\txst \frac12} (E_+\pm
E_-)^2-{\txst \frac12}\, (\vec{v\,}\vec{q\,})^2-\Delta^2\, .
\nonumber
 \ee
 Then after the separation of the pole part we obtain
\be
(G_{+}\!\cdot\!G^h_{-})&=& a^2\,\rho\,
\frac{\om^2-(\vec{v\,}\vec{q\,})^2-2\, \Delta^2}{2\, \Delta^2}\,
\frac12\, g_T(\vec{n\,},\om,\vec{q\, })
\nonumber\\ &+&
a^2\,\rho\,\frac12\intop_{-\infty}^{+\infty}\rmd \epsilon_p \Big[
\frac{n(E_-)-n(E_+)}{2\,E_+\, E_-}\, (E_- -E_+)
+\frac{1-n(E_-)-n(E_+)}{2\,E_+\, E_-}\, (E_++E_-)\Big]\,
\nonumber\\
&=& a^2\,\rho\,
\frac{\om^2-(\vec{v\,}\vec{q\,})^2-2\, \Delta^2}{2\, \Delta^2}\,
\frac12\, g_T(\vec{n\,},\om,\vec{q\, })+
a^2\,\rho\intop_{-\infty}^{+\infty}\rmd \epsilon_p \frac{1-2\, n(E_p)}{2\, E_p}\, .
\nonumber
 \ee
The last divergent term must be cut of at some scale $\xi\sim \mu$ and is nothing else but the
quantity $A_0$ in Eq. (\ref{A0}), which enters the gap equation (\ref{gapeq}). The relation
(\ref{Legrel3}) follows now immediately.

Now we are in the position to  generalize expressions for the loop
functions (\ref{depend}) for $T\neq 0$:
\be
M(\vec{n\,},\om,\vec{q\,}) &=&
-a^2\,\rho\,\frac{\om+\vec{q}\,\vec{v}}{2\, \Delta}\,g_T(\vec{n\,},\om,\vec{q\,})\,,
\nonumber\\
N(\vec{n\,},\om,\vec{q\,}) &=&
a^2\, \rho\, \frac{\om^2-(\vec{q}\,\vec{v})^2}{4\, \Delta^2}\,g_T(\vec{n\,},\om,\vec{q\,})\,,
\nonumber\\
O(\vec{n\,},\om,\vec{q\,};P) &=&
a^2\, \rho\, \Big[ \frac{\om+\vec{q}\,\vec{v}}{4\, \Delta}+ \frac{\om-\vec{q}\,\vec{v}}{4\, \Delta}\, P \Big]\,
g_T(\vec{n\,},\om,\vec{q\,})\,
\label{loopsT}
 \ee
with $g_T$ given by Eq.~(\ref{gT1}). The $L$ function is less straightforward.
We find
 \be
&&L=(G_{+}\cdot G_{-}-P\,F_+F_-)
\nonumber\\
&&=-a^2\,
\rho\intop_{-\infty}^{+\infty}\frac{\rmd \epsilon_p}{2\,E_+\,E_-}\,
\Big[
\frac{n(E_-)-n(E_+)}{\om^2 - (E_+ - E_-)^2}\,
(E_+ - E_-)
\Big(
\Delta^2\, (1+P) +\big((\vec{v\, }\vec{q\,})^2-(E_-+ E_+)^2\big)
\frac{\om +(\vec{v\,}\vec{q\,})}{2\, (\vec{v\, }\vec{q})}
\Big)
\nonumber\\
&&\qquad\qquad
-\frac{1- n(E_-)- n(E_+)}{\om^2 - (E_+ + E_-)^2}\, (E_+ + E_-)
\Big(
\Delta^2\, (1+P) +\big((\vec{v\, }\vec{q\,})^2-(E_-- E_+)^2\big)
\frac{\om +(\vec{v\,}\vec{q\,})}{2\, (\vec{v\, }\vec{q})}
\Big)
\Big]
\nonumber\\
&&=-a^2\, \rho\,\frac{1+P}2\, g_T
- a^2\,\rho\,\frac{\om +(\vec{v\,}\vec{q\,})}{2\, (\vec{v\, }\vec{q})}
\intop_{-\infty}^{+\infty}\frac{\rmd \epsilon_p}{2\,E_+\,E_-}\,
\Big[
\frac{\big(n(E_-)-n(E_+)\big)}{\om^2 - (E_+ - E_-)^2}\,
(E_+ - E_-) \big((\vec{v\, }\vec{q\,})^2-(E_-+ E_+)^2\big)
\nonumber\\
&&\qquad\qquad
-\frac{\big(1- n(E_-)- n(E_+)\big)}{\om^2 - (E_+ + E_-)^2}\,
(E_+ + E_-)
\big((\vec{v\, }\vec{q\,})^2-(E_-- E_+)^2\big)
\Big]\, .
 \ee
Then with the help of the relation
 \be
(\vec{v\, }\vec{q\,})^2-(E_- \pm E_+)^2=\frac{4\,
(\vec{v\, }\vec{q\,})^2\, \Delta^2}{(E_- \mp E_+)^2-(\vec{v\, }\vec{q\,})^2}
 \ee
we present
\be
\frac{(\vec{v\, }\vec{q\,})^2-(E_- \pm E_+)^2}{\om^2 - (E_+ \mp E_-)^2} &=&
\frac{4\, (\vec{v\, }\vec{q\,})^2\, \Delta^2}{\om^2-(\vec{v\, }\vec{q\,})^2}\,
\Bigg(\frac{1}{\om^2 - (E_+ \mp E_-)^2}
-
\frac{1}{(\vec{v\, }\vec{q\,})^2-(E_- \mp E_+)^2} \Bigg) .
\label{magic}
 \ee
Finally after comparison with Eq.~(\ref{gT1}) we derive
 \be
L(\vec{n\,},\om,\vec{q\,};P)=a^2\, \rho\, \Big[
\frac{\vec{v}\, \vec{q}}{\om- \vec{v}\,\vec{q}}\,
\big( g_T(\vec{n\,},(\vec{v\,}\vec{q\,}),\vec{q\,}) - g_T(\vec{n\,},\om,\vec{q\,})\big)
-\frac{1+P}{2}\,g_T(\vec{n\,},\om,\vec{q\,})\Big]\, .
\label{Lfun}
 \ee
For $T=0$, $g_T(\vec{n\,},(\vec{v\,}\vec{q\,}),\vec{q\,})=1$ and the old result (\ref{loopsT0}) is
recovered.

The relation between the function $\kappa$ used in the Leggett equation (\ref{Leggett:kappa}) and
the functions used in the Larkin-Migdal equations (\ref{depend}) is derived with the help of  the
substitution of Eq.~(\ref{magic}) into Eq.~(\ref{GhGh+GG}) and by making use of the fact that the
function $g_T$ is even function of $\vec{q}$. Thus we obtain
 \be
\kappa(\vec{n\,},\om,\vec{q\,})&=& a^2\, \rho\,
\frac{(\vec{v}\, \vec{q\,})^2}{\om^2- (\vec{v}\,\vec{q\,})^2}\,
\big( g_T(\vec{n\,},(\vec{v\,}\vec{q\,}),\vec{q\,}) - g_T(\vec{n\,},\om,\vec{q\,})\big)
\nonumber\\
&=& \frac{\vec{v}\, \vec{q}}{\om+ \vec{v}\,\vec{q}}\, L(\vec{n\,},\om,\vec{q\,};-1)
={\txst\frac12}\big( L(\vec{n\,},\om,\vec{q\,};-1)+L(\vec{n\,},-\om,\vec{q\,};-1)\big)\,.
 \ee
 From (\ref{Lfun}) and (\ref{gT1}) we verify that
$L(\vec{n\,},-\om,-\vec{q\,};P)=L(\vec{n\,},\om,\vec{q\,};P)$ and
$L(-\vec{n\,},\om,-\vec{q\,};P)=L(\vec{n\,},\om,\vec{q\,};P)$, and
therefore
\be
\langle\kappa(\vec{n\,},\om,\vec{q\,})\rangle_{\vec{n}} &=&
{\txst\frac12}
\big(
 \langle L(\vec{n\,},\om,\vec{q\,};-1) \rangle_{\vec{n}}
+\langle L(\vec{n\,},-\om,\vec{q\,};-1)\rangle_{\vec{n}}
\big)
\nonumber\\ &=& {\txst\frac12}
\big(
 \langle L(\vec{n\,},\om,\vec{q\,};-1) \rangle_{\vec{n}}
+\langle L(-\vec{n\,},\om,\vec{q\,};-1)\rangle_{\vec{n}}
\big) \nonumber\\
&=& \langle L(\vec{n\,},\om,\vec{q\,};-1) \rangle_{\vec{n}}\, .
\label{kappa-L}
 \ee

\section{Current conservation}\label{app:Ward}

Here we prove the transversality of the polarization tensor (\ref{def:Pi}), $\Pi^{\mu\nu}\,
q_\nu=0$ which guarantees the conservation of the vector current in a superfluid Fermi
liquid.

As we have shown in Appendix~\ref{app:loop}, the loop functions $M$, $N$, $O$ and $L$ are
expressed through the only  $g_T$ function defined for arbitrary temperature. From  definitions
(\ref{curleL}), (\ref{curleLtilde}) of  functions ${\mathcal{L}}_a$ and
$\widetilde{\mathcal{L}}_a$ and  expressions for the loop functions (\ref{loopsT}) and
(\ref{Lfun}) we deduce that
 \be
\om\,\langle \widetilde{\mathcal{L}}(\vec{n},q;P_{a,0})
\rangle_{\vec{n}} =
\langle\widetilde{\mathcal{L}}(\vec{n},q;P_{a,1})\,(\vec{v\,}\vec{q}\,)\rangle_{\vec{n}}\,.
\label{eqLL}
 \ee
From Eq. (\ref{def:Pi}) we obtain
 \be
\Pi_V^{0\mu}(\om,\vec{q\,})\, q_\mu &=&\frac{e_V}{a}\,
\om \, \langle \chi_{V,0}(\vec{n\,},q) \rangle_{\vec{n}}
-\langle \vec{q\,} \vec{\chi}_{V,1}(\vec{n},q)\rangle_{\vec{n}}\,,
\label{wardid:1}\\
\Pi^{i\mu}_V(\om,\vec{q\,})\, q_\mu &=& \frac{e_V}{a}\,
 \big\langle \vec{v\,\,}_i\,\big(\om \chi_{V,0}(\vec{n\,},q)-\vec{q}\,\vec{\chi}_{V,1}(\vec{n\,},q)\big)
 \big\rangle_{\vec{n}}\,,
\label{wardid:2} \ee
with $e_V=1$\,. To demonstrate  transversality of the polarization tensor for the  vector current
we need to prove vanishing of these components.

Applying the averages given in Eqs.~(\ref{aver:chi0},\ref{aver:chi1q}), Eq.~(\ref{eqLL}) and using
that $P_{a,0}=-P_{a,1}$ we obtain
 \be
&&\langle \om \chi_{a,0}(\vec{n},q)  - \vec{q\,}\vec{\chi}_{a,1}(\vec{n},q)\rangle_{\vec{n}}
\nonumber\\
&&\qquad = \om\,\gamma_a(q;P_{a,0})\, \langle \mathcal{L}(\vec{n},q;P_{a,0}) \rangle_{\vec{n}}\, \tau_{a,0}^\om -
\gamma_a(q;-P_{a,1})\,
\langle \widetilde{\mathcal{L}}(\vec{n},q;P_{a,1})\,(\vec{n}\cdot\vec{q}\,)\rangle_{\vec{n}}\,
\tau_{a,1}^\om
\nonumber\\
&&\qquad =\tau_{a,0}^\om\gamma_a(q;P_{a,0})\,\Big(\om\, \langle
\widetilde{\mathcal{L}}(\vec{n},q;P_{a,0}) \rangle- \langle
\widetilde{\mathcal{L}}(\vec{n},q;P_{a,1})\,(\vec{v}\cdot\vec{q}\,)\rangle\Big)=0\,.
\nonumber
 \ee
Remarkably, this relation holds both for the vector current and for the axial-vector currents, i.e.
$\Pi_a^{0\mu}(\om,\vec{q\,})\, q_\mu =0$.

Now we turn to the second necessary relation~(\ref{wardid:2}). Calculating the
product $\vec{q}\,\vec{\chi}_{V,1}$ we use Eq.~(\ref{chivec:sol}).
Consider first
\be
\vec{q}\,\delta\vec{\chi}_{V,1}(\vec{n},q) &=& e_V a\,\rho\,
\frac{(\om+\vec{v}\,\vec{q}\,)\, g_T(\vec{n},q)}
 {\langle (\om^2-(\vec{v}\,\vec{q}\,')^2)\, g_T(\vec{n},q)\rangle_{\vec{n'}}}\,
\langle (\vec{v\,}\vec{q\,})^2\, g_T(\vec{n},q) \rangle_{\vec{n'}}
\nonumber\\
&+&\gamma_V(q;+1)\,\Gamma_V^\om\,\Big(
\mathcal{L}_V(\vec{n},q;+1)\,
 \langle\widetilde{\mathcal{L}}_V(\vec{n'}, q; -1) (\vec{q}\,\vec{n\,}')\rangle_{\vec{n}'}
-\mathcal{L}_V(\vec{n},q;-1)\,(\vec{q}\,\vec{n})\,
\langle\widetilde{\mathcal{L}}_V(\vec{n},q;+1)\rangle_{\vec{n}}
\Big)\, \tau_{V,1}^\om\,.
\nonumber
 \ee
Here we used Eqs. (\ref{loopsT}). Making use  of this relation and Eqs.~(\ref{idL}) and (\ref{eqLL}) we calculate
\be
&&\big\langle (\vec{v}\,\vec{q}\,)\,
\big(\om \chi_{V,0}-\vec{q}\, \vec{\chi}_{V,1}\big)
\big\rangle_{\vec{n}}
\nonumber\\
&&\qquad=\big\langle (\vec{v}\,\vec{q}\,)\,
\big[\om\,\gamma_{V}(q;+1)\,\mathcal{L}_V(\vec{n},q;+1)
-(\vec{v}\,\vec{q}\,)\,\gamma_{V}(q;+1)\,\mathcal{L}_V(\vec{n},q;-1)
- \vec{q}\,\delta\vec{\chi}_{V,1}(\vec{n},q) \big]
\big\rangle_{\vec{n}}\, \tau_{V,0}^\om
\nonumber\\
&&\qquad=
  \big(\langle(\vec{v}\,\vec{q}\,)\mathcal{L}(\vec{n},q;+1)\rangle_{\vec{n}}\,\om
 -\big\langle (\vec{v}\,\vec{q}\,)^2\,\mathcal{L}(\vec{n},q;-1)\big\rangle_{\vec{n}}\big)\,\,
\tau_{V,0}^\om
\nonumber\\
&&\qquad\qquad
- a\,\rho\,\frac{\langle(\vec{v}\,\vec{q}\,)\,(\om+\vec{q}\,\vec{v}\,)\, g_T(\vec{n},q)\rangle_{\vec{n}}}
{ \langle (\om^2-(\vec{v}\,\vec{q\,})^2)\, g_T(\vec{n},q)\rangle_{\vec{n}}}
 \langle (\vec{v}\,\vec{q\,})^2\, g_T(\vec{n},q) \rangle_{\vec{n}} .
 \ee
 Interestingly, there is no dependence  on $\Gamma_0^\om$
in the last expression. Finally, using explicit expression for the
loop functions (\ref{loopsT},\ref{Lfun}) we arrive at the identity
\be
\big\langle \vec{v\,}_i\,\big(\om
\chi_{V,0}(\vec{n},q)-\vec{q}\,\vec{\chi}_{V,1}(\vec{n},q)\big)
\big\rangle_{\vec{n}}=e_V a\,
\rho\,\langle\vec{v\,}_i(\vec{v}\,\vec{q}\,)\rangle_{\vec{n}}\, ,
 \ee
 which means that $\Pi^{i\mu}_V(\om,\vec{q\,})\, q_\mu =0$.



\begin{thebibliography}{99}
\bibitem{FRS76} G.~Flowers, M.~Ruderman and P.G.~Sutherland, Ap. J. {\bf 205}, 541 (1976).
\bibitem{VS87} D.N.~Voskresensky and A.V.~Senatorov, Sov. J. Nucl. Phys. {\bf 45}, 411 (1987).
\bibitem{SV87} A.V.~Senatorov and D.N.~Voskresensky, Phys. Lett. B {\bf 184}, 119 (1987).
\bibitem{MSTV90} A.B.~Migdal, E.E.~Saperstein, M.A.~Troitsky and D.N.~Voskresensky, Phys. Rept. {\bf 192}, 179 (1990).
\bibitem{SVSWW97} Ch.~Schaab, D.~Voskresensky, A.D.~Sedrakian, F.~Weber and  M.K.~Weigel, Astron. Astrophys. {\bf 321}, 591 (1997).
\bibitem{Minimal} D.~Page, astro-ph/9802171; D.~Page, J.M.~Lattimer, M.~Prakash and
A.W.~Steiner, ArXive: astro-ph/0403657.
\bibitem{KHY} A.D.~Kaminker, P.~Haensel and D.G.~Yakovlev, Astron. Astrophys. {\bf 345}, L14 (1999).
\bibitem{YKL99}  D.G.~Yakovlev, A.D.~Kaminker and K.P.~Levenfish, Astron. Astrophys. {\bf{343}}, 650 (1999).
\bibitem{YLS99}D.G.~Yakovlev, A.D.~Kaminker, O.Y.~Gnedin and P.~Haensel, Phys. Rept. {\bf 354}, 1
(2001).
\bibitem{V01} D.N.~Voskresensky, Lect. Notes Phys. {\bf 578},  467 (2001); ArXive: astro-ph/0101514.
\bibitem{BGV04} D.~Blaschke, H.~Grigorian and D.N.~Voskresensky, Astron. Astrophys. {\bf 424}, 979 (2004).
\bibitem{GV05} H.~Grigorian and D.N.~Voskresensky, Astron.Astrophys. {\bf 444}, 913 (2005).
\bibitem{PGW} D.~Page, U.~Geppert and F.~Weber, Nucl. Phys. A {\bf 777}, 497 (2006).
\bibitem{Sedr07} A.~Sedrakian, Prog. Part. Nucl. Phys. {\bf 58}, 168 (2007).
\bibitem{KR} J.~Kundu and S.~Reddy, Phys. Rev. C {\bf 70}, 055803 (2004).
\bibitem{LP} L.B.~Leinson and A.~Perez, Phys. Lett. B {\bf 638}, 114 (2006); ArXive: astro-ph/0606653.
\bibitem{SMS} A.~Sedrakian, H.~M\"uther and P.~Schuck, Phys. Rev. C {\bf 76}, 055805 (2007);
A.~Sedrakian and J.~Keller,  arXiv:1001.0395 [nucl-th].
\bibitem{KV08} E.E.~Kolomeitsev and D.N.~Voskresensky, Phys. Rev. C {\bf 77}, 065808 (2008), arXiv:0802.1404 [nucl-th].
\bibitem{L08} L.B.~Leinson,  Phys. Rev. {\bf C78}, 015502 (2008), arXiv:0804.0841 [astro-ph].
\bibitem{SR09} A.W.~Steiner and S.~Reddy, Phys. Rev. {\bf C79}, 015802 (2009).
\bibitem{PLPS}
D. Page, J.M. Lattimer, M. Prakash and A.W. Steiner, Astrophys. J.
{\bf  707}, 1131 (2009).
\bibitem{LeinsonWrong} L.B.~Leinson, Phys. Rev. {\bf C79}, 045502 (2009).

\bibitem{Nambu} Y.~Nambu, Phys. Rev. {\bf 117}, 648 (1960).
\bibitem{Schriffer} J.R.~Schriffer, ``Theory of Superconductivity'', Benjamin, N.Y., 1964.
\bibitem{LM63} A.I.~Larkin and A.B.~Migdal, Sov. Phys. JETP {\bf 17}, 1146 (1963).
\bibitem{M67} A.B.~Migdal, ``Theory of Finite Fermi Systems and Properties of
Atomic Nuclei'', Willey and Sons, N.Y. 1967; second. ed. (in Rus.), Nauka, Moscow, 1983.
\bibitem{Leg65a} A.J.~Leggett, Phys. Rev.  {\bf 140}, A1869 (1965).
\bibitem{L66}    A.J.~Leggett, Phys. Rev. {\bf 147}, 119 (1966).
\bibitem{KV95} J.~Knoll and  D.N.~Voskresensky, Phys. Lett. B {\bf 351}, 43 (1995);
Ann. Phys. (N.Y.) {\bf 249}, 532 (1996).
\bibitem{VGL61} V.G.~Vaks, V.M.~Galitskii and  A.I.~Larkin, Sov. Phys. JETP {\bf 14}, 1177 (1961).
\bibitem{KVfuture} E.E.~Kolomeitsev and D.N.~Voskresensky, in preparation.
\bibitem{Migdal63} A.B.~Migdal, Sov. Phys. JETP {\bf 16}, 1366 (1963).
\bibitem{Migdal59} A.B.~Migdal, Nucl. Phys. {\bf 13}, 655 (1959).
\bibitem{ML64} A.B.~Migdal and A.I.~Larkin, Nucl. Phys. {\bf 51}, 561 (1964).
\bibitem{Migdal64} A.B.~Migdal, Nucl. Phys. {\bf 57}, 29 (1964).
\bibitem{W93} H.A.~Weldon,  Phys. Rev. D {\bf 47}, 594 (1993).  



\end{thebibliography}
\end{document}